\documentclass[aps,a4paper,amsmath,amssymb,twocolumn,
floatfix,groupedaddress]{revtex4-1}
\usepackage{bm,amsmath,amssymb,amsfonts}
\usepackage[dvips]{graphicx}
\usepackage{color}
\usepackage{xcolor}
\definecolor{dgreen}{rgb}{0,0.6,0}
\definecolor{brown}{rgb}{0.9,0.4,0.3}
\usepackage{hyperref}

\hypersetup{
    pdfnewwindow=true,      % links in new window
    colorlinks=true,       % false: boxed links; true: colored links
    linkcolor=magenta,          % color of internal links
    citecolor=blue,        % color of links to bibliography
    filecolor=blue,      % color of file links
    urlcolor=blue        % color of external links
}
\newcommand\bea{\begin{eqnarray}}
\newcommand\eea{\end{eqnarray}}
\newcommand\beq{\begin{equation}}
\newcommand\eeq{\end{equation}}

\newcommand{\ie}{{\it i.\,e.}, }

\newcommand{\viz}{{\it viz.}, } 

\begin{document}

% Title
\title{A fractal geometry immersed in a hierarchical magnetic flux distribution}
%%%%%%%%%%%%%%%%%%%%%
\author{Biplab Pal} 
%\email{biplabpal2008@gmail.com}
\email[E-mail: ]{biplab@nagalanduniversity.ac.in}
%%%%%%%%%%%%%%%%%%%%%
\affiliation{Department of Physics, School of Sciences, 
Nagaland University, Lumami-798627, Nagaland, India}
%%%%%%%%%%%%%%%%%%%%%
\date{\today}
%%%%%%%%%%%%%%%%%%%%%
\begin{abstract}
Fractal geometry presents us with a self-similarity in their pattern 
at various length scales that is prevalent in our natural world. 
We present theoretical model of a Sierpinski gasket (SPG) fractal 
geometry with a deterministic perturbation in the form of a hierarchical 
distribution of magnetic flux. Such flux configuration induces a 
deterministic disorder in the Aharonov-Bohm (AB) phases picked up by the 
electron wavefunction. Using the tight-binding formalism, we show that 
by tunning the strength of the hierarchy parameter of those AB phases, 
one can systematically engineer quantum states in a SPG fractal 
lattice. In addition to this, we have also observed that by controlling 
the strength of this hierarchy parameter in the magnetic flux, one can 
effectively regulate the persistent current in the SPG fractal 
structure. This characteristic is found to be true for various 
filling factors. Our results could be useful for designing 
nanoelectronic devices using molecular fractal structures fabricated 
by chemical synthesis technique. 
\end{abstract}
%%%%%%%%%%%%%%%%%%%%%
\maketitle

%%%%%%%%%%%%%%%%%%%%%%%%%%%%%%%%%%%%%%%%%%%%%%%%%%%%%%%%%%%%%%%%%%%%%%%
\section{Introduction}
\label{sec:intro}
The terminology of fractal geometries which was initially emerged 
purely as a mathematical concept~\cite{mandelbrot-book}, has 
generated intense research interest in various domains of condensed 
matter physics and material science in recent times. 
Although deterministic fractal geometries belongs to the class of 
aperiodic systems, they are not completely random; they possess 
certain self-similar beautiful pattern and symmetrical structure 
in their construction. This often helps theoreticians to analytically 
investigate various interesting properties appearing in such fractal 
structures. Also, deterministic growth rule allows one to grow such 
fractal structures in the laboratory using the state-of-the-art 
self-assembly or lithography techniques~\cite{self-assembly-1,
self-assembly-2,hexagonal-SPG}, and perform fascinating experiments 
with them. 

One of the key reasons why fractal lattice structures have drawn a 
lot of research attention of late is due to the fact that they 
possess non-integer Hausdorff dimension $D_f$. This has ignited a 
lot of curiosity in the mind of the researchers and led to a number of 
intriguing investigations in this direction, viz., 
topological character of electronic states on 
fractals~\cite{titus,shreya}, flat bands in fractal-like 
lattices~\cite{biplab,atanu,hahafi-fractal-FB,chen-fractal-FB-1,
chen-fractal-FB-2,sougata-fractal-FB}, Hall conductivity~\cite{askar-Hall,
cm-smith-Hall} and possible existence of anyons in fractal 
structures~\cite{biplab-PRR-2020}, are to name a few of them. 
The rapid advancement in experimental tools and 
fabrication methods has also played a major role in developing renewed 
interest in the physics of fractal geometries. These technological 
progress has enabled us to synthesize tailor-made fractal structures in 
the laboratory using various techniques, such as, self-assembled 
molecular growth of fractal materials~\cite{self-assembly-1,
self-assembly-2}, designing quantum fractal using atomic 
manipulation in a scanning tunnelling 
microscope~\cite{cm-smith-nat-phys}, 
fractal photonic lattice consisting of waveguides realized using 
femtosecond-laser-writing technology~\cite{m-segev} etc.
  
Inspired by these experimental advancements in creating the fractal 
structures in reality, various interesting theoretical studies 
concerning the deterministic fractal structures have taken place in 
recent times. This includes works on energy level statistics in fractal 
clusters~\cite{yorikawa-JPC2019,askar-PRB2019}, confinement 
of electrons in fractal plane using electric field~\cite{yuan-PRB2020}, 
electronic fractal pattern in building molecular fractal 
systems~\cite{latge-pccp2022}, role of anisotropy on magneto-transport 
in fractal network~\cite{maiti-PLA2020}, Josephson effect in a fractal 
geometry~\cite{vladimir-APL2024}, and stability of loop current states in 
fractal structures built with Bose-Einstein condensates~\cite{anna-PRA2024} 
among others. 
%######################################################
\begin{figure}[ht]
\includegraphics[clip,width=0.8\columnwidth]{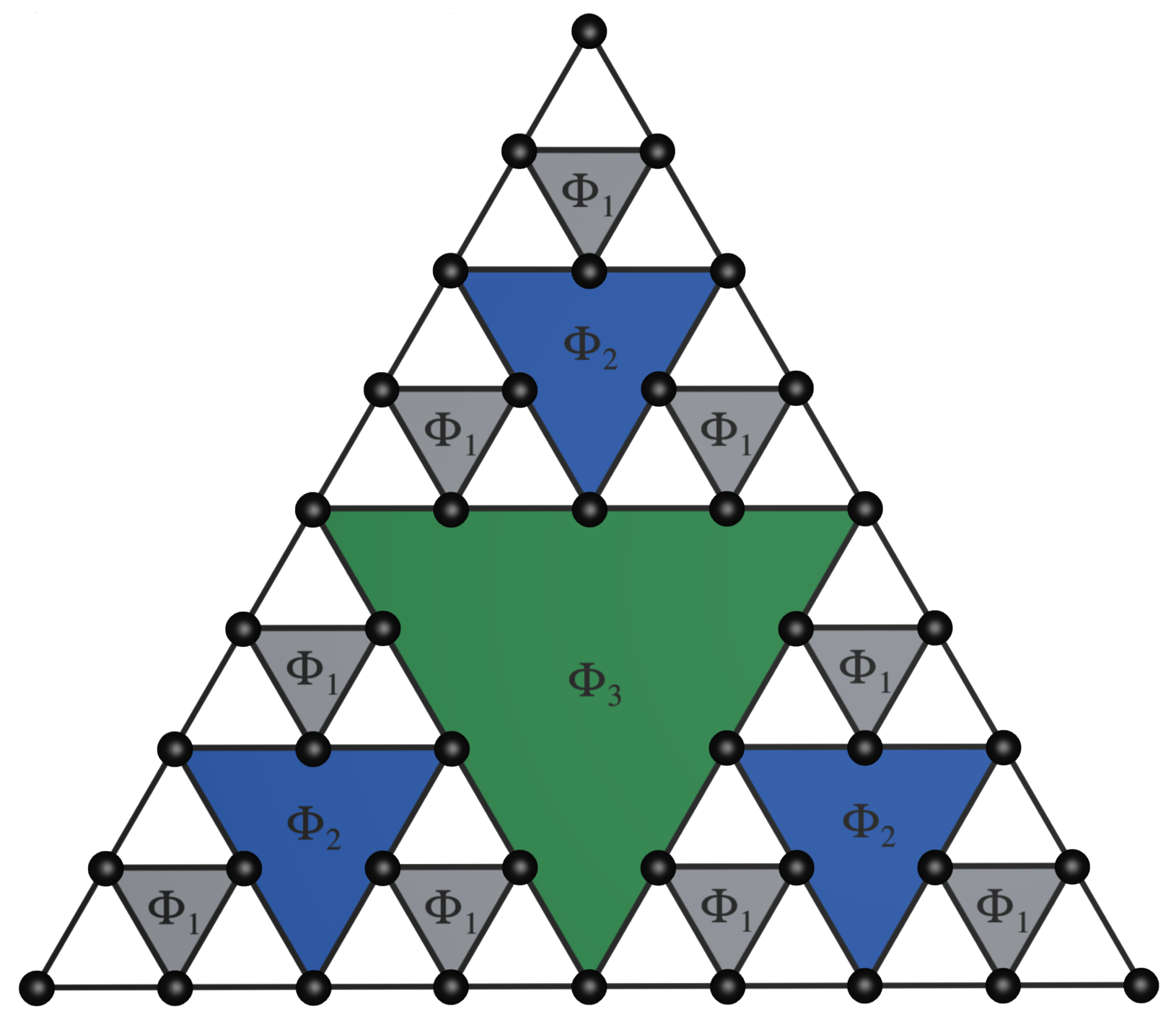}
\caption{Schematic diagram of a Sierpinski gasket fractal 
lattice with hierarchical distribution of a magnetic flux following 
the pattern $\Phi_{\ell}=\lambda^{\ell}\Phi$, where $\ell$ represents 
the generation index.}
\label{fig:lattice}
\end{figure}
%######################################################  
It is expected that, in future, further advancement in nanotechnology 
and fabrication techniques will enable us to do more delicate experiments 
with these fractal lattice models. Such possibility inspires us to 
undertake this theoretical investigation presented in this article.   

We consider a Sierpinski gasket (SPG) fractal lattice structure immersed 
in a deterministic hierarchical distribution of magnetic fluxes. 
The ramification of a SPG fractal structure is such that, it offers us 
with down-headed triangular loops of gradually increasing size (See 
Fig.~\ref{fig:lattice}). Following this ramification pattern, we introduce 
a hierarchical distribution of magnetic fluxes 
$\Phi_{\ell}=\lambda^{\ell}\Phi$ ($\lambda$ being the hierarchy parameter) 
through these down-headed triangular closed loops. Such distribution of 
the magnetic fluxes will be manifested as a distribution of 
Aharonov-Bohm (AB) phases, whenever an electron will hop from one 
atomic site to another atomic site along these down-headed triangular 
closed loops. This provides us with an interesting situation, where we 
have a deterministic disorder in the distribution of the AB phases in 
this closed-loop fractal geometry controlled by a hierarchy parameter 
$\lambda$. It is to be noted that, in Fig.~\ref{fig:lattice}, 
although schematically we have shown that the fluxes are embedded in the 
downward triangles, but in reality, along the arms of the upward triangles, 
we will also have various distribution of the Aharonov-Bohm phases depending 
on their position in the fractal geometry for any given finite generation. This 
means that, the upward triangles are also containing some effective fluxes. One 
may realize such a complex distribution of magnetic fluxes by putting up some 
synthetic gauge fields~\cite{gardin-arXiv2024} in various downward plaquettes, 
which may be controlled locally in each plaquette.       

In this work, we investigate the effect of this usual distribution of 
magnetic fluxes on the energy eigenvalue spectrum of the SPG fractal 
network as a function of the hierarchy parameter $\lambda$. The presence 
of the magnetic flux through such closed-loop fractal lattice geometry 
induces a circulating current in these systems known as persistent 
current~\cite{maiti-PRB2010}. It is a pertinent question to ask, 
may we have some mechanism by which we can control the persistent 
current in such closed-loop fractal lattice geometries? In this study, 
we address this question and show that, it is indeed possible to 
effectively control the magnitude of the persistent current in this 
SPG fractal lattice by tunning the hierarchy parameter $\lambda$ in 
the distribution of the magnetic flux. Our model may be suitable for 
mesoscopic or nanoscopic experiments using fractal structures. 
In what follows, we present the schematic and mathematical framework 
of our model in Sec.~\ref{sec:model}. We discuss the effect of the 
hierarchy parameter on the energy eigenvalue spectrum of the SPG 
fractal lattice model in Sec.~\ref{sec:energy-spectrum}. The role of 
the hierarchy parameter in controlling the persistent current in such 
complex systems is elaborated in Sec.~\ref{sec:persistent-current}. 
Finally, in Sec.~\ref{sec:summary}, we draw our conclusion highlighting 
the summary of our findings and their utility, and the scope of future 
study in this direction.       
%%%%%%%%%%%%%%%%%%%%%%%%%%%%%%%%%%%%%%%%%%%%%%%%%%%%%%%%%%%%%%%%%%%%%%%
\section{The model and the mathematical framework}
\label{sec:model}
We consider a theoretical model of a SPG fractal lattice 
which follows certain deterministic ramification pattern in its 
growth rule as shown in Fig.~\ref{fig:lattice}. The number of sites in a 
$\ell$-th generation system is given by the following formula: 
%-----------------------------------------------------------------------------
\begin{equation}
\mathcal{N}(\ell) = \dfrac{3}{2}\big(1+R^{\ell}\big),
\label{eq:growth-rule} 
\end{equation}
%-----------------------------------------------------------------------------
where $R$ is the repetition factor (which is equal to 3 for this 
particular fractal model) and $\ell$ denotes the generation index. The 
single-electron states in such a lattice structure can be described by 
the following Hamiltonian within the tight-binding framework:  
%-----------------------------------------------------------------------------
\begin{equation}
\bm{\mathcal{H}} = \sum_{n} \epsilon_{n}c_{n}^{\dagger}c_{n} + 
\sum_{\langle m,n\rangle} \big(t_{mn}e^{i\theta_{\ell}}c_{m}^{\dagger}c_{n} 
+ \textrm{h.c.}\big),
\label{eq:hamiltonian}
\end{equation}
%-----------------------------------------------------------------------------
where first term represents the on-site energy and the second term 
represents the nearest-neighbor hopping integral. $\theta_{\ell} = 
\dfrac{2\pi\Phi_{\ell}}{3(2^{\ell-1})\Phi_{0}}$ is the Aharonov-Bohm 
phase~\cite{A-B-Phase} associated with the nearest-neighbor hopping parameter 
$t_{mn}$, whenever an electron hops along a bond in the down-headed triangular 
loops pierced by an external magnetic flux $\Phi_{\ell}=\lambda^{\ell}\Phi$, 
where $\lambda$ is the hierarchy parameter. 
$\Phi$ is measured in units of fundamental flux quantum $\Phi_{0}=hc/e$. 
$c_{n}^{\dagger}\,(c_{n})$ denotes the creation (annihilation) operator for 
an electron at a site $n$, and $\epsilon_{n}$ is the on-site potential for 
the $n$-th atomic site. 

Following the Eq.~\eqref{eq:hamiltonian}, we can construct the Hamiltonian 
matrix for this fractal lattice model of different system sizes, which will 
allow us to extract the information about the energy spectrum as well as 
for the persistent current of this system as a function of the deterministic 
disorder distribution in the AB phases. It is to be noted that, 
because of the self-similar characteristics of a fractal system, once we get 
some interesting properties in such systems for a small system size, it 
generally retains those properties even when we go to a large system size. 
Thus, considering the possibility of experimental realization of this model 
with such an unusual hierarchical distribution of magnetic fluxes in 
the real-life laboratory, we consider moderately large system for our 
theoretical calculations. At first, we study the role of the hierarchy 
parameter $\lambda$ which controls the hierarchical distribution of the 
magnetic flux $\Phi_{\ell}$ on the energy eigenvalue spectrum for this model. 
This is elaborated in the next section.  
%%%%%%%%%%%%%%%%%%%%%%%%%%%%%%%%%%%%%%%%%%%%%%%%%%%%%%%%%%%%%%%%%%%%%%%
\section{Effect of the hierarchy parameter on the energy eigenvalue spectrum}
\label{sec:energy-spectrum} 
\begin{figure}[ht]
\includegraphics[clip,width=0.49\columnwidth]{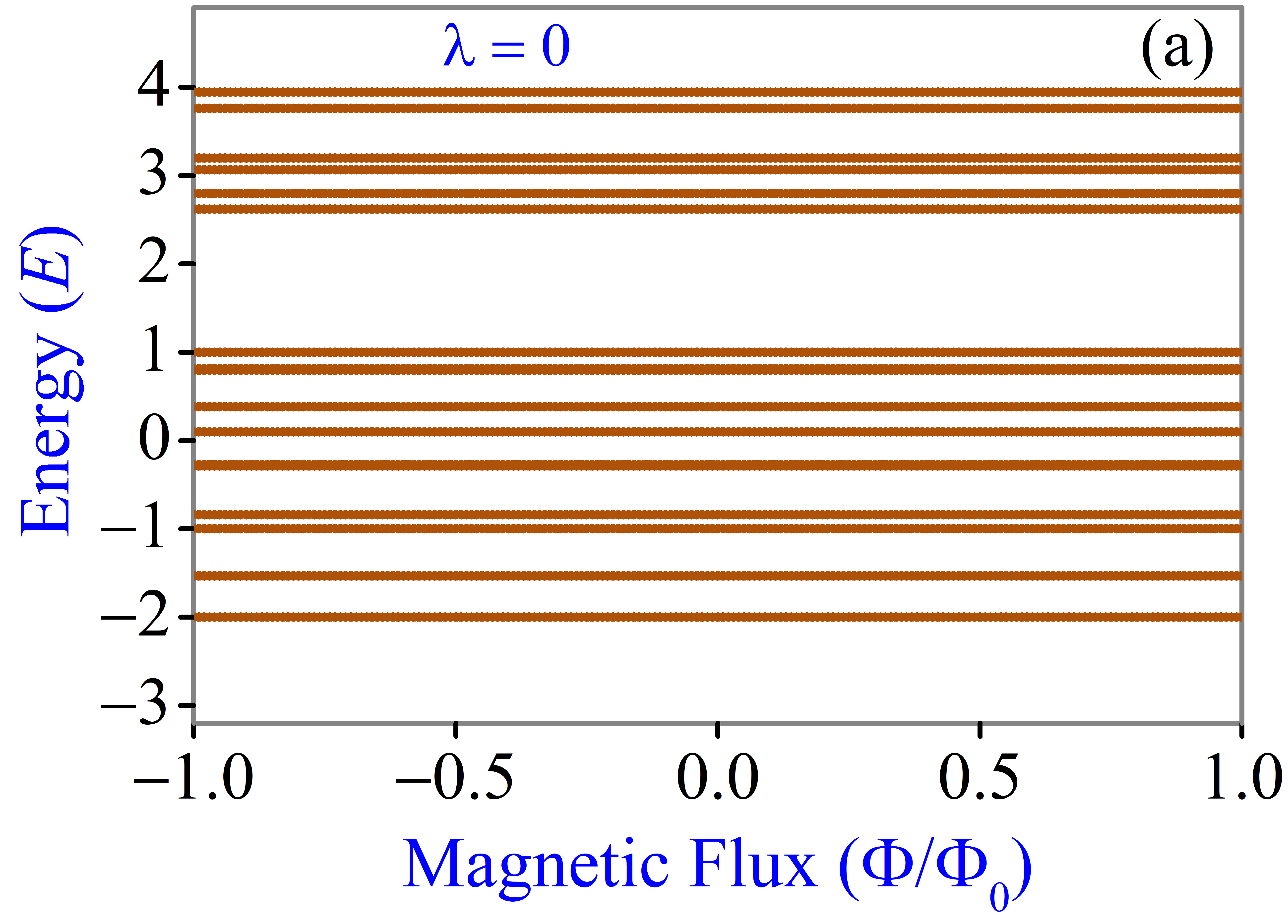}
\includegraphics[clip,width=0.49\columnwidth]{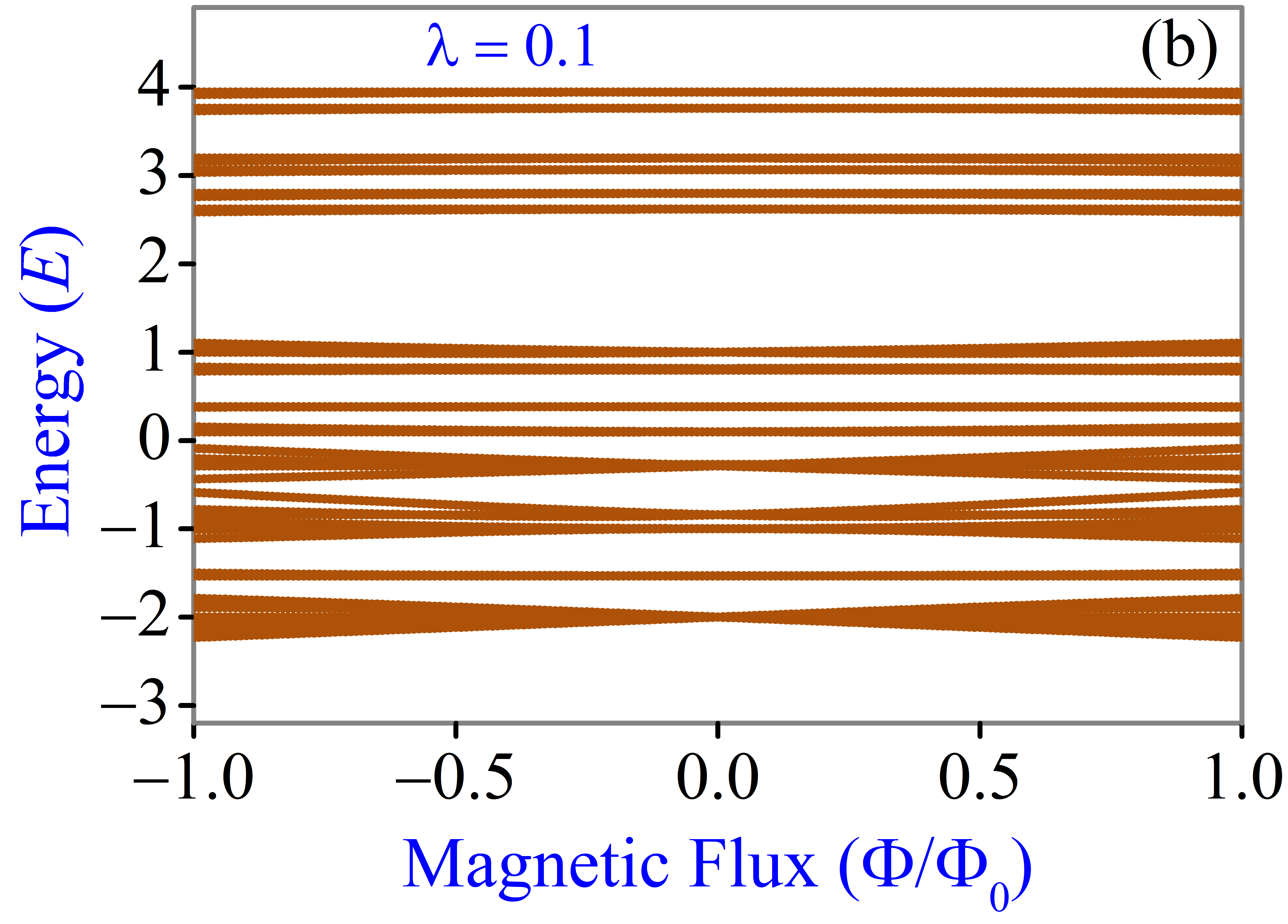}
\vskip 0.2cm
\includegraphics[clip,width=0.49\columnwidth]{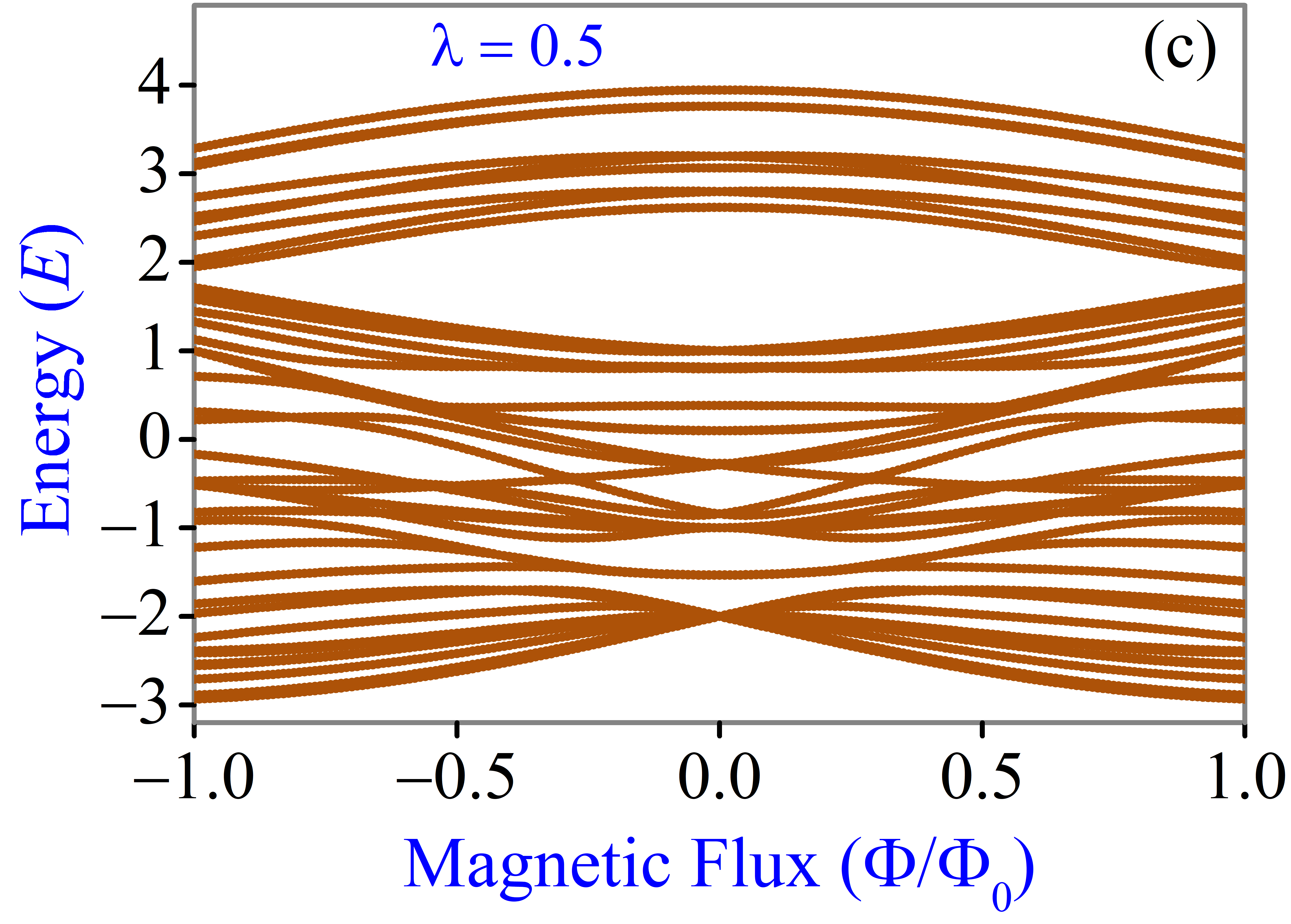}
\includegraphics[clip,width=0.49\columnwidth]{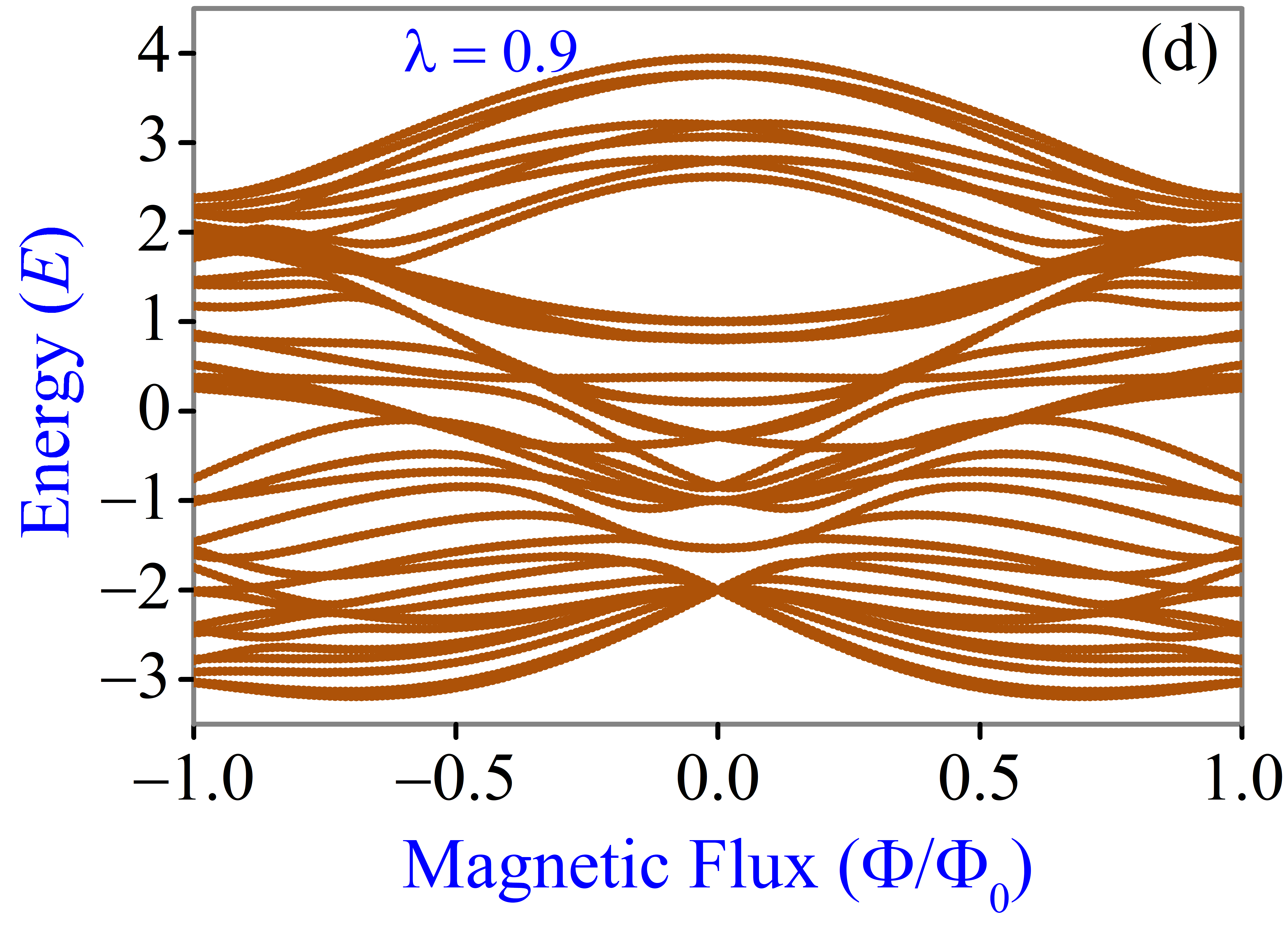}
\caption{Variation of the energy eigenvalue spectrum for 
a 3rd generation SPG lattice structure (with $\mathcal{N}(3) = 42$) 
as a function of the hierarchical magnetic flux for different 
values of the hierarchy parameter, \viz (a) $\lambda=0$, 
(b) $\lambda=0.1$, (c) $\lambda=0.5$, and (d) $\lambda=0.9$.}
\label{fig:eng-spec-gen3}
\end{figure}

\begin{figure}[ht]
\includegraphics[clip,width=0.49\columnwidth]{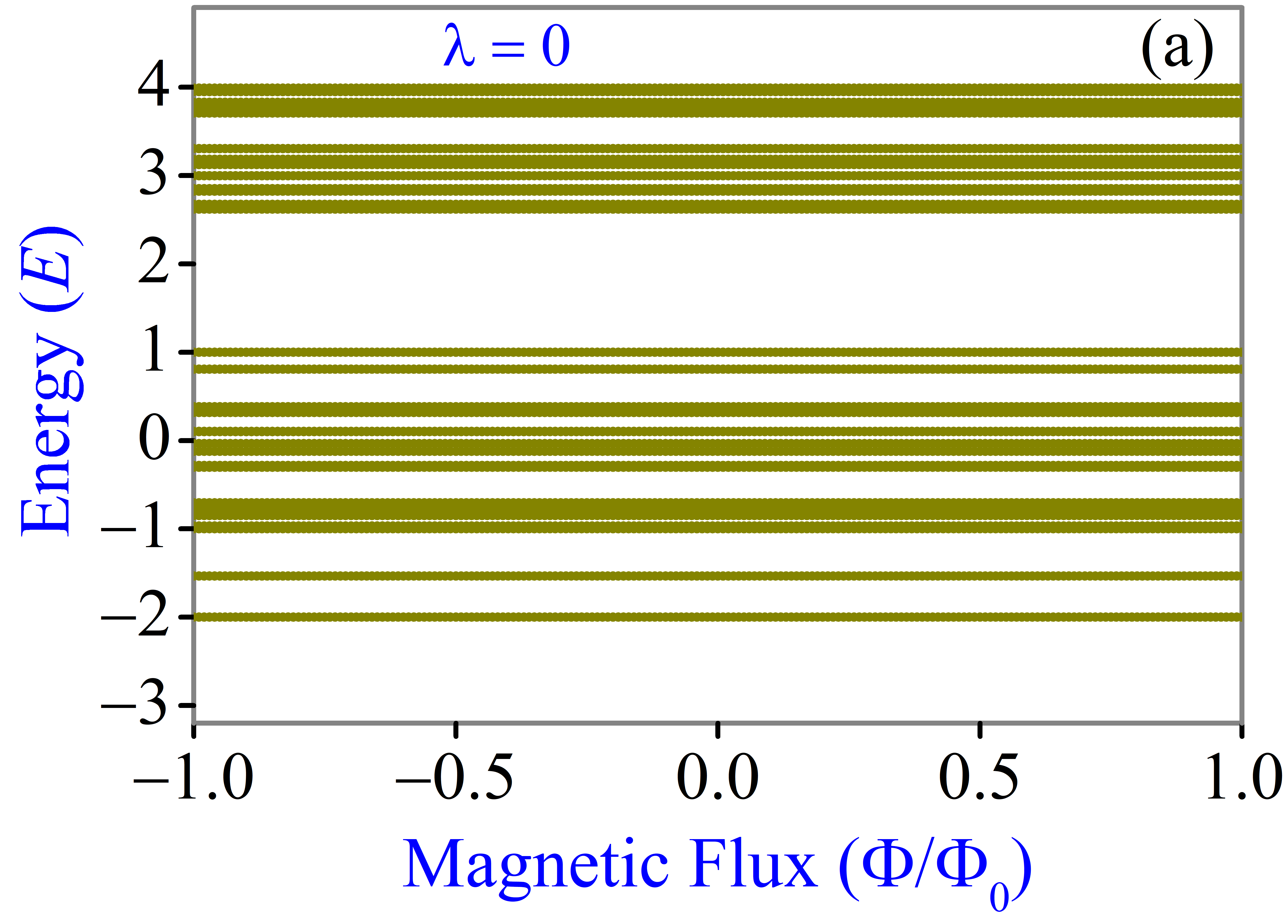}
\includegraphics[clip,width=0.49\columnwidth]{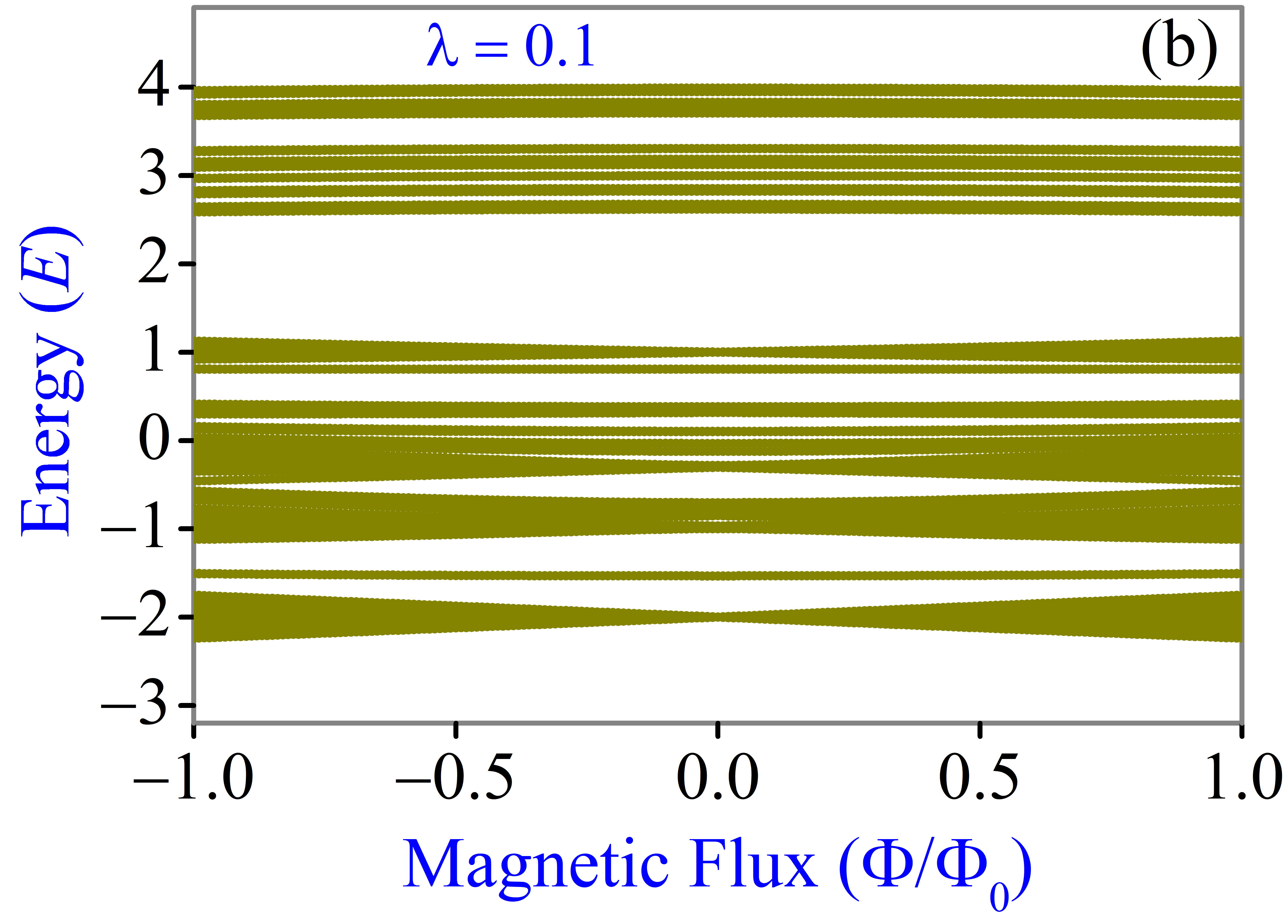}
\vskip 0.2cm
\includegraphics[clip,width=0.49\columnwidth]{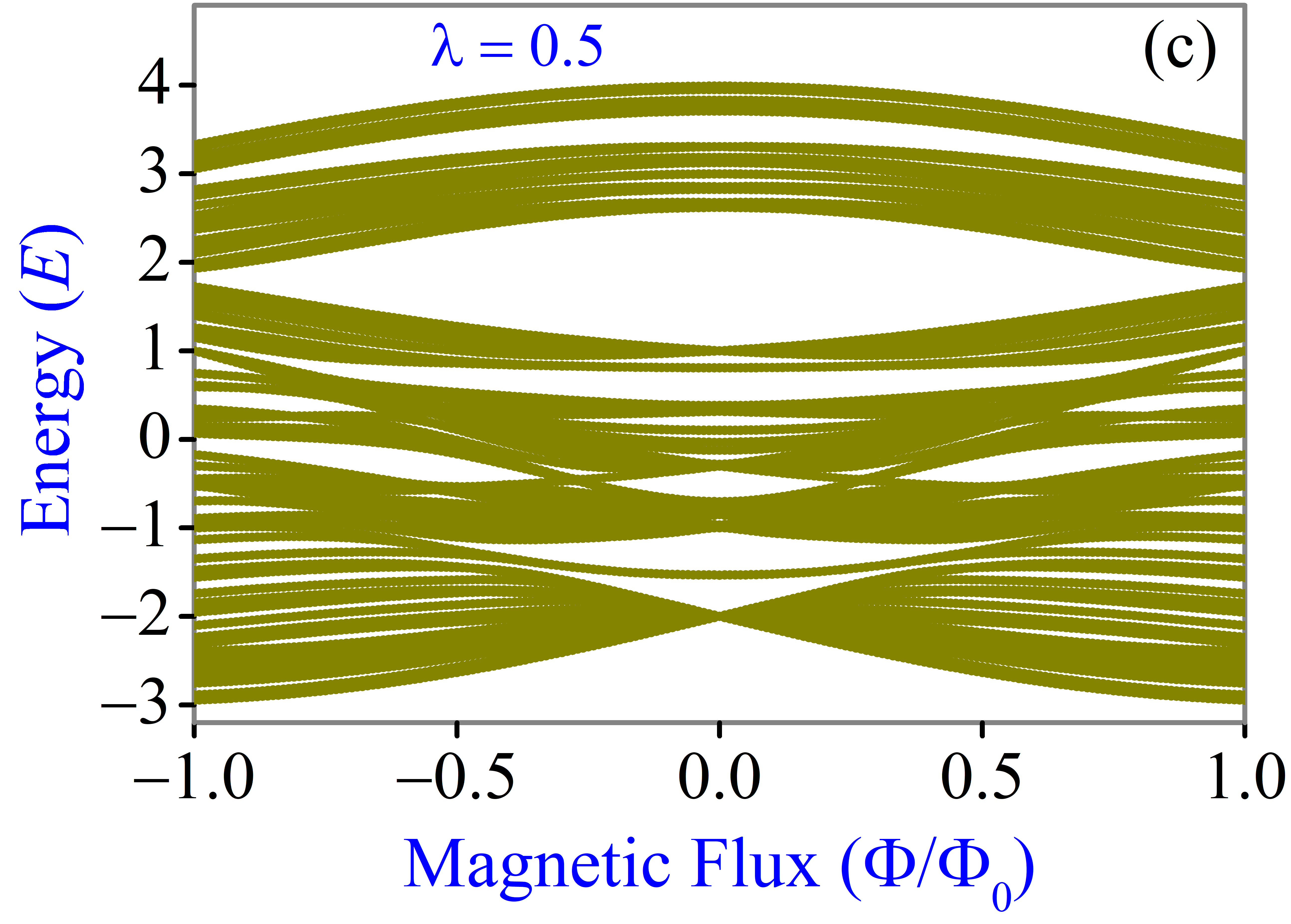}
\includegraphics[clip,width=0.49\columnwidth]{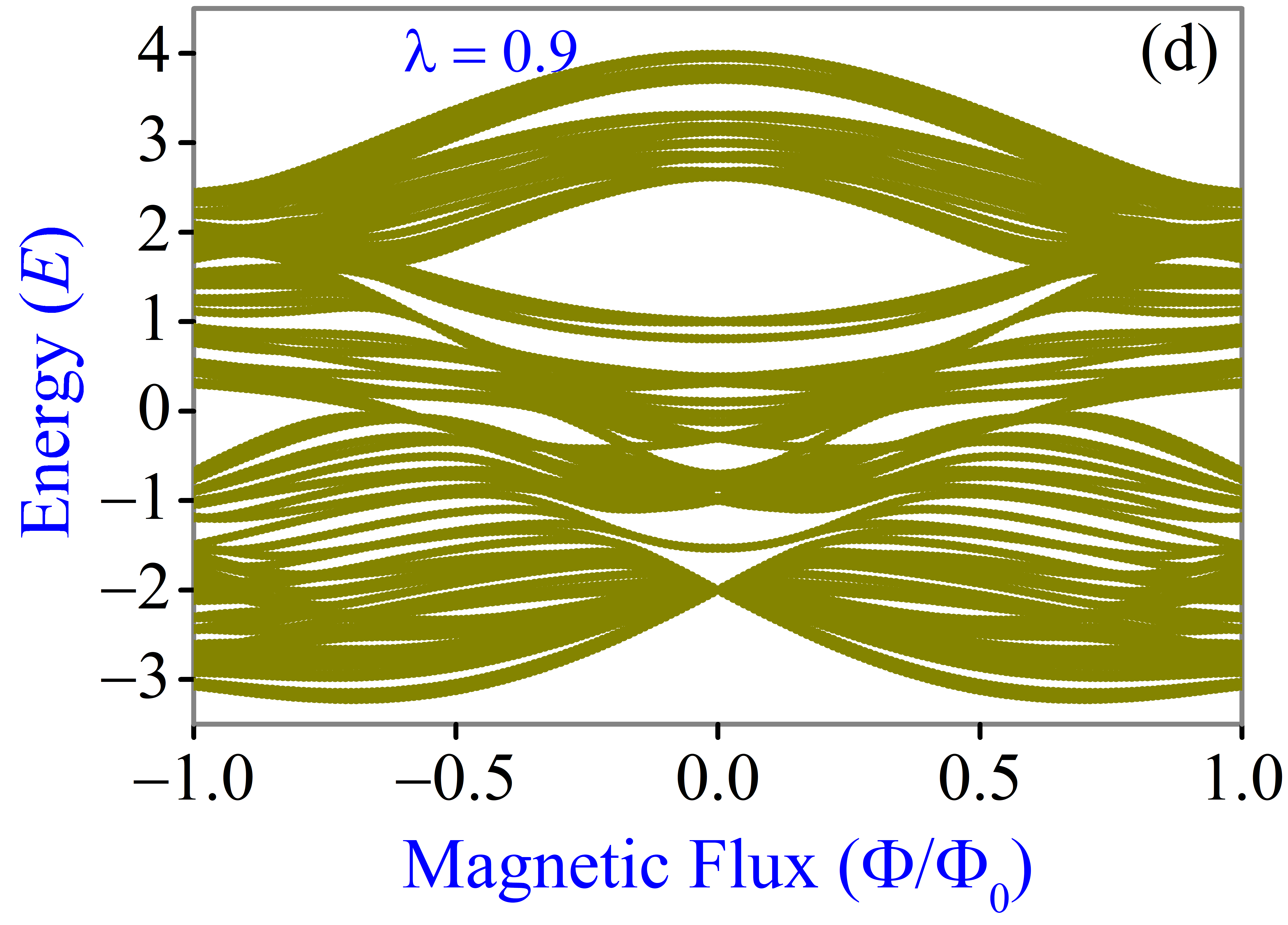}
\caption{Energy eigenvalue distribution for a 4th 
generation SPG lattice structure (with $\mathcal{N}(4) = 123$) 
as a function of the hierarchical magnetic flux with 
values of the hierarchy parameter as (a) $\lambda=0$, 
(b) $\lambda=0.1$, (c) $\lambda=0.5$, and (d) $\lambda=0.9$.}
\label{fig:eng-spec-gen4}
\end{figure}

In this section, we discuss the role of the hierarchy parameter $\lambda$ 
on the energy eigenvalue spectrum of the SPG lattice. We have first 
explicitly constructed the Hamiltonian matrix for the SPG lattice using 
Eq.~\eqref{eq:hamiltonian}. We take the on-site potential $\epsilon_{n}=0$ 
for all the sites, and consider only nearest-neighbor constant coupling 
between the sites $t_{mn}=t=1$. The direction of the hierarchical fluxes 
$\Phi_{\ell}$ (where $\ell=1,2,3,\hdots$) in different down-headed triangular 
closed plaquettes is taken to be in the out-of-plane direction. 
Accordingly, the electrons will pick up the AB phases $\pm \theta_{\ell}$ 
(`$+$' is for forward hopping and `$-$' is for backward hopping). Once the 
Hamiltonian matrices are constructed for different generations of the SPG 
lattice, one can easily diagonalize them to evaluate the energy eigenvalue 
spectrum for the lattice structure. Here, we present the results for the 
3rd and the 4th generation SPG lattice structure in 
Fig.~\ref{fig:eng-spec-gen3} and Fig.~\ref{fig:eng-spec-gen4}, respectively. 
We note that, the energy ($E$) of the particle is measured 
in units of the hopping integral $t$.
%######################################################
%###################################################### 

We start our discussion with $\lambda=0$, \ie in absence of the hierarchy 
parameter. It is clear that for $\lambda=0$, the hierarchical magnetic flux 
distribution in all the plaquettes vanishes. Hence, under this condition, 
we get the energy eigenvalue distribution for the SPG lattice geometry 
in absence of the magnetic flux. The energy eigenvalue spectrum exhibit 
highly degenerate, discrete, fractal characteristics 
[see Fig.~\ref{fig:eng-spec-gen3}(a)]. Then, we incorporate a small 
non-zero value of $\lambda=0.1$. With this small non-zero value of $\lambda$, 
we instantly observe that, the degeneracy in the energy spectrum is slightly 
lifted [Fig.~\ref{fig:eng-spec-gen3}(b)]. As we gradually tune the strength of 
the hierarchy parameter $\lambda$ more, the energy eigenvalue spectrum become 
increasingly non-degenerate 
[Fig.~\ref{fig:eng-spec-gen3}(c) and~\ref{fig:eng-spec-gen3}(d)]. Thus, by 
controlling the strength of the hierarchy parameter $\lambda$, one can 
systematically control the degeneracy of the energy spectrum in a SPG fractal 
structure. 
%######################################################
%###################################################### 

It is to be noted that, the self-similar aspect of a fractal structure is, 
in general, reflected in all its properties. Therefore, one can expect that, 
our observations appearing in Fig.~\ref{fig:eng-spec-gen3} for a 3rd 
generation SPG fractal structure will qualitatively persist even when we 
go to a higher generation fractal structure. To confirm this fact, we have 
also evaluated the energy spectrum for a 4th generation SPG fractal with 
different values of the hierarchy parameter $\lambda$. This is depicted in 
Fig.~\ref{fig:eng-spec-gen4}(a)-(d). It is clearly evident that, the 
qualitative structure of the energy spectrum remains same as in 
Fig.~\ref{fig:eng-spec-gen3}(a)-(d), apart from appearance of some 
additional energy eigenvalues which appear in the spectrum due to 
increase in the system size. We anticipate the similar feature to be 
sustained even for larger SPG fractal structures. 
It is worth to point out that, the ramification of the SPG fractal structure 
is such that one cannot have such hierarchical distribution of magnetic 
fluxes in the upward triangular plaquettes (see Fig.~\ref{fig:lattice}). 
In upward triangular plaquettes, one can only have a uniform distribution of 
the magnetic flux as discussed in Ref.~\cite{maiti-PRB2010,chakrabarti-PRB2005}. 
%######################################################
\begin{figure}[ht]
\includegraphics[clip,width=0.49\columnwidth]{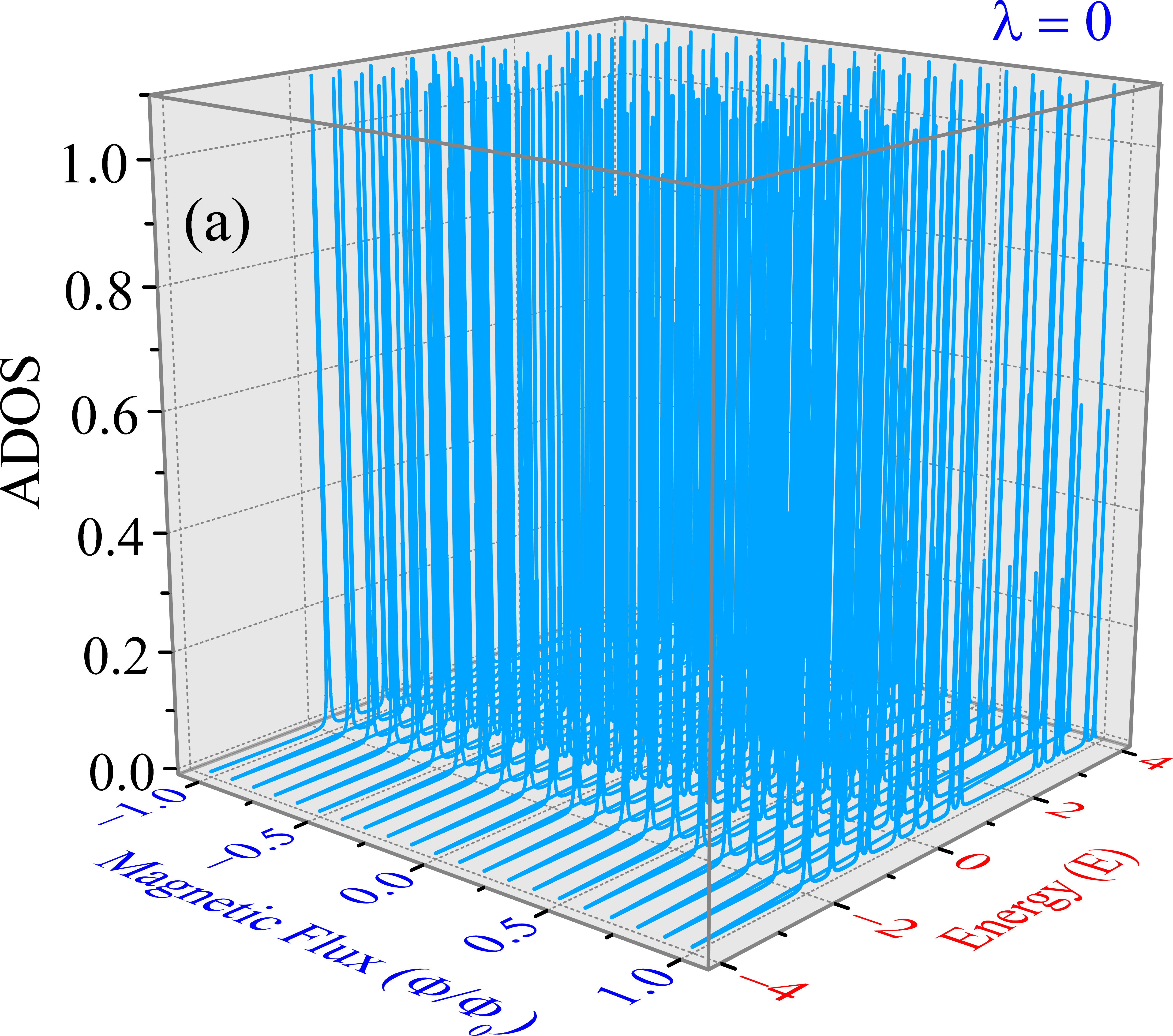}
\includegraphics[clip,width=0.49\columnwidth]{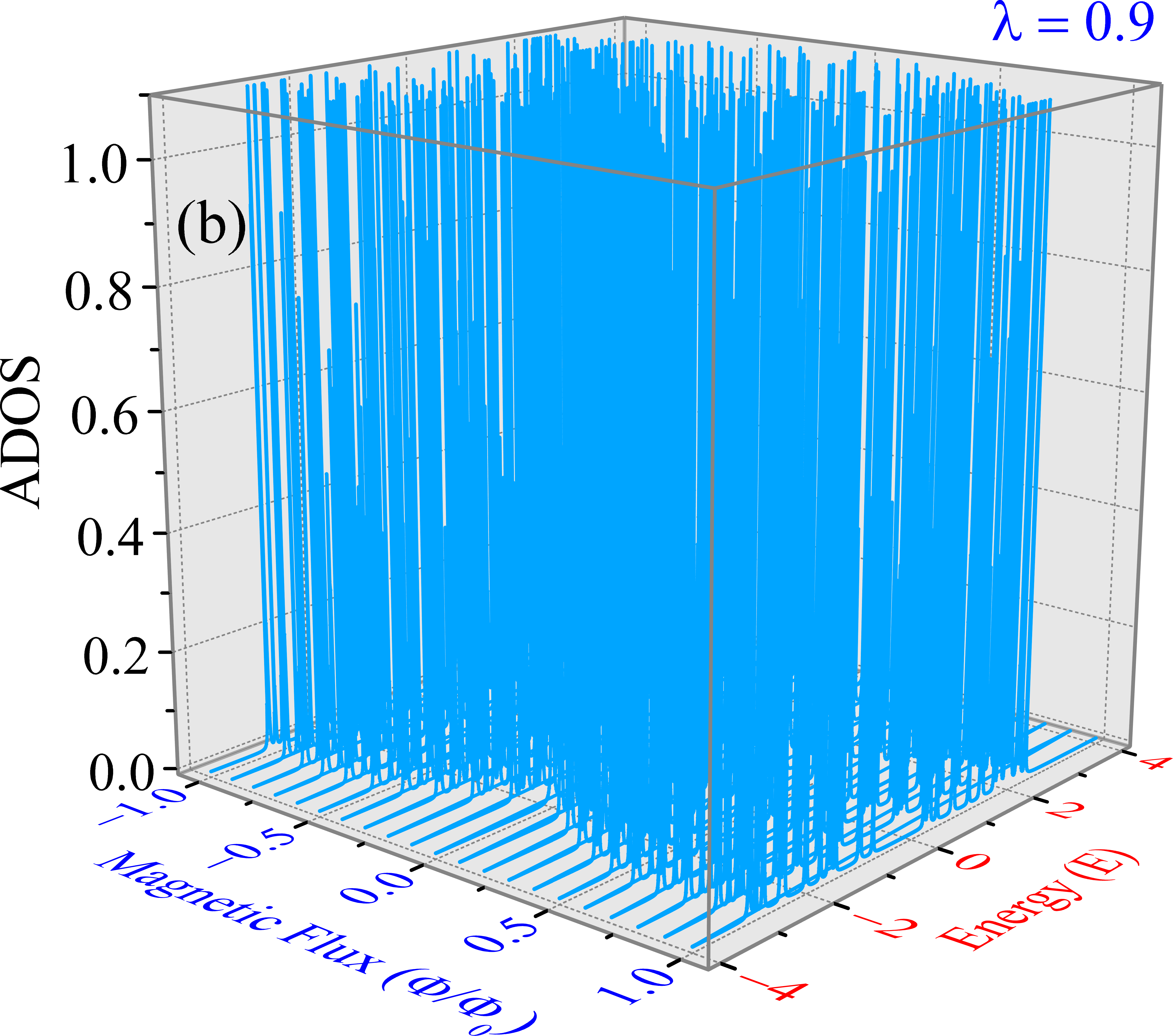}
\caption{Average density of states (ADOS) for a 3rd 
generation SPG fractal structure as a function of the 
energy and hierarchical magnetic flux. 
(a) is for $\lambda=0$ and (b) is for 
a non-zero value of $\lambda=0.9$.}
\label{fig:dos}
\end{figure}
%###################################################### 
%
%######################################################
\begin{figure*}[ht]
\includegraphics[clip,width=0.6\columnwidth]{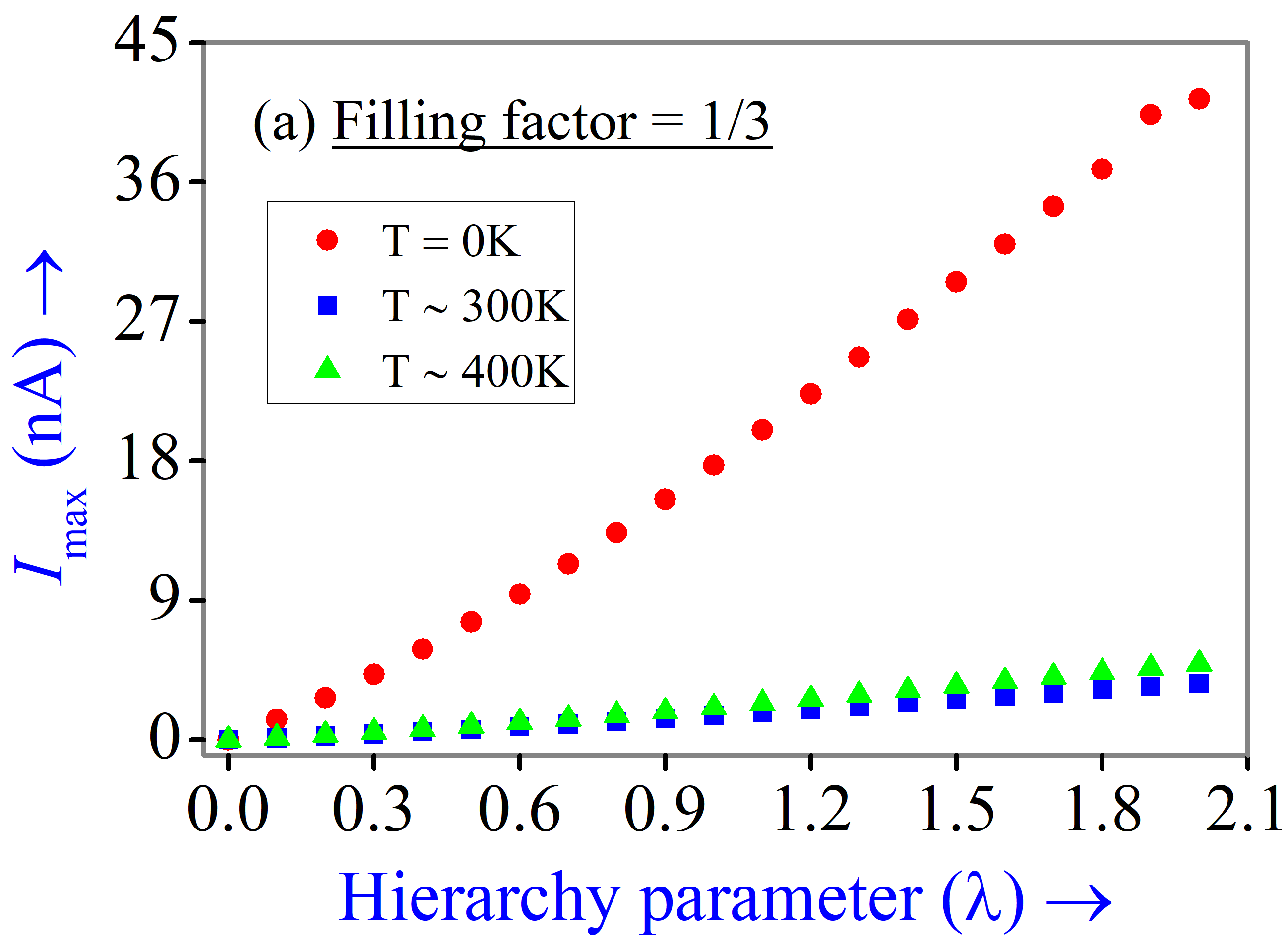}
\includegraphics[clip,width=0.6\columnwidth]{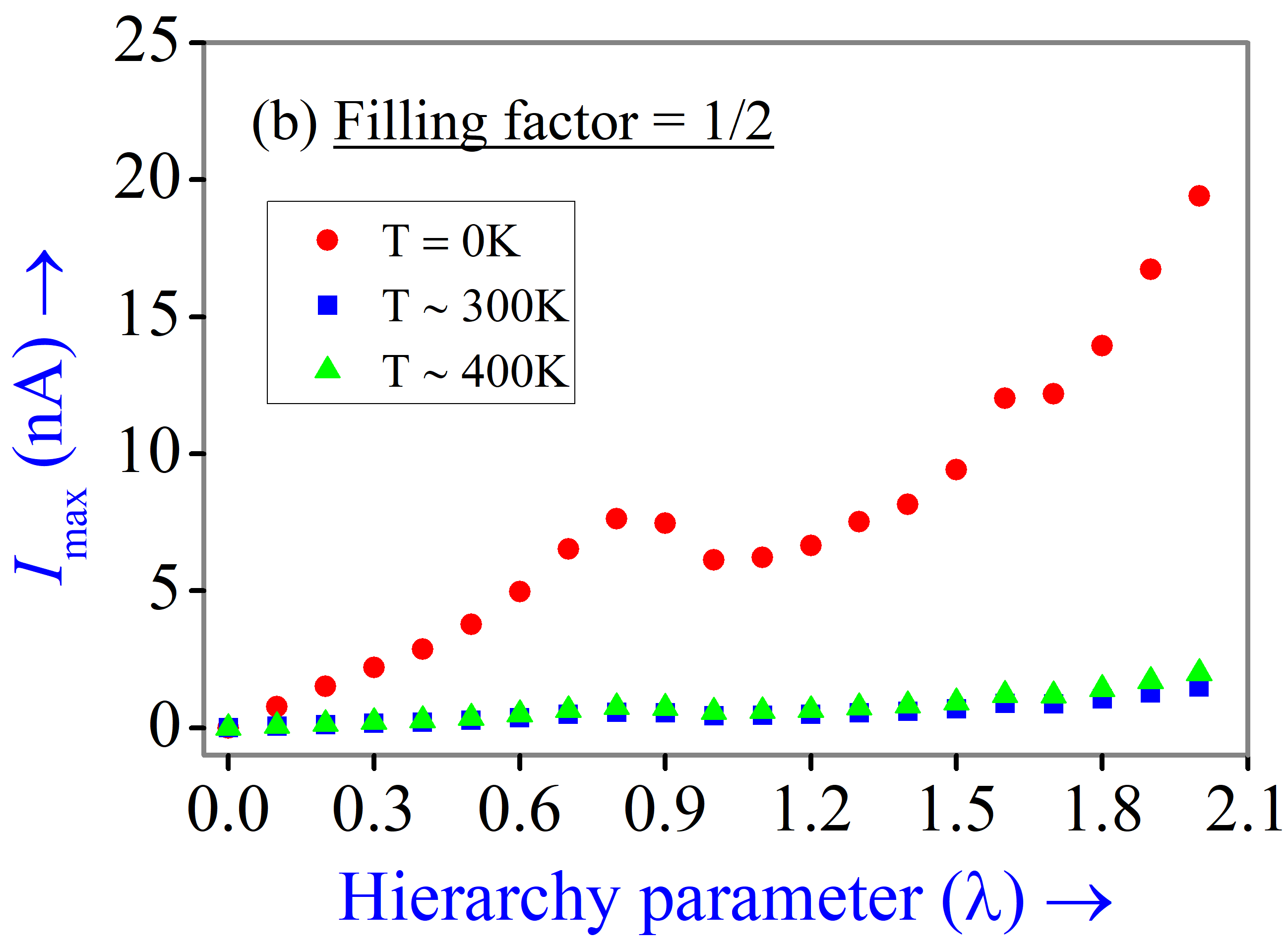}
\includegraphics[clip,width=0.6\columnwidth]{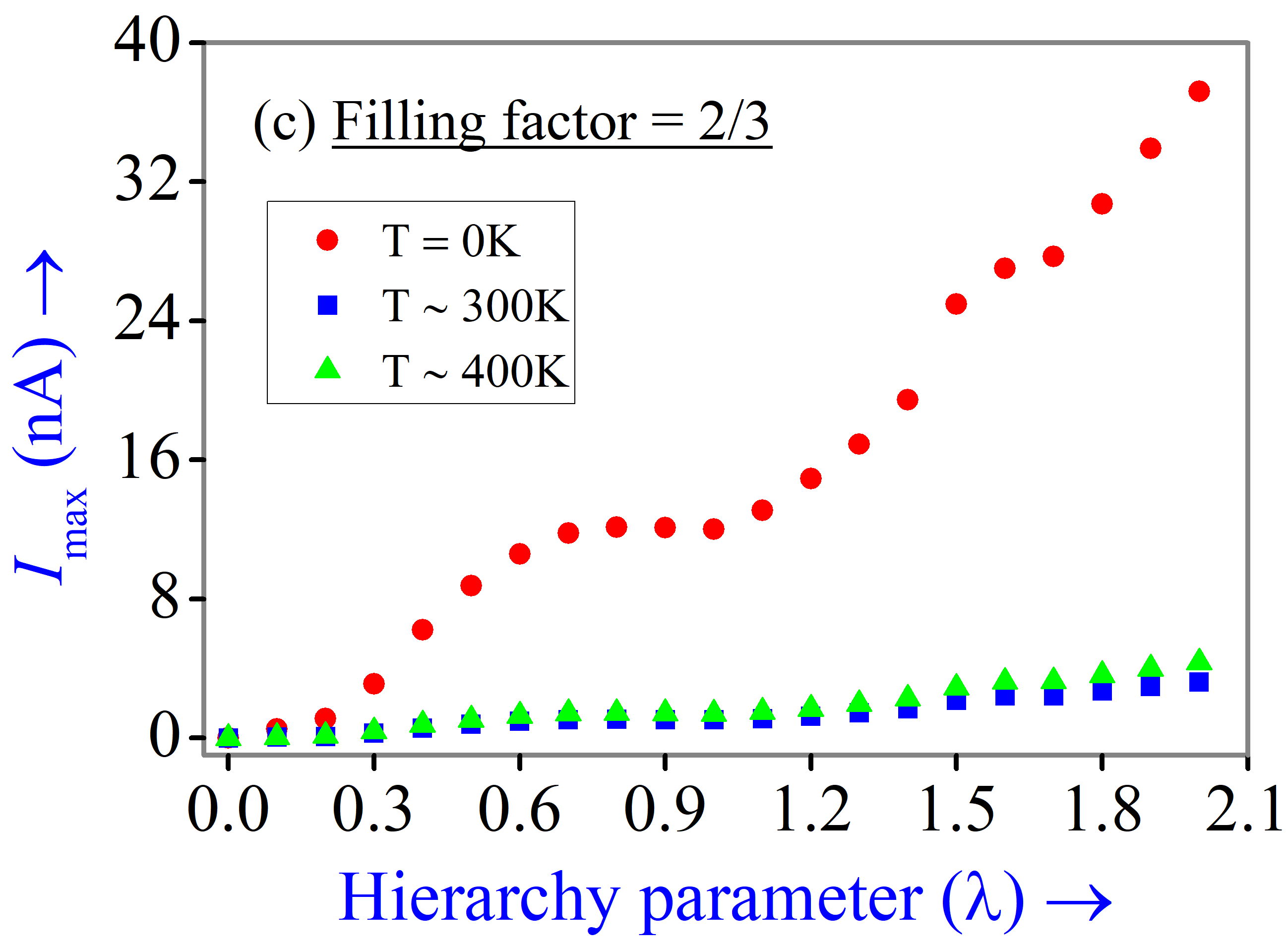}
\caption{Variation of the persistent current 
($I_{\textrm{max}}$) as a function of the hierarchy parameter 
$\lambda$ for a 3rd generation SPG fractal lattice for 
the temperature $T = 0$ and $T > 0$, respectively.  
(a) is for a filling factor $\nu=1/3$, (b) is for $\nu=1/2$, 
and (c) represents $\nu=2/3$.}
\label{fig:persistent-current-gen3}
\end{figure*}
%#####################################################

To corroborate our results on the energy spectrum of the SPG fractal lattice 
as a function of the hierarchy parameter, we have calculated the average 
density of states (ADOS) for a finite generation SPG fractal structure. 
To compute the ADOS, we have used the Green's function technique which 
people uses in general for such fractal lattice 
models~\cite{nandy-JPCM2015,pal-PRB2012,pal-PRA2013,pal-EPJB2012}. 
Using the Green's function technique, ADOS can be defined as,
%-----------------------------------------------------------------------------
\begin{equation}
\rho(E,\Phi) = -\dfrac{1}{\mathcal{N}\pi}\textrm{Im} 
\Big[ \textrm{Tr}\left[\bm{\mathcal{G}}(E,\Phi) \right]\Big],
\label{eq:ADOS}
\end{equation} 
%-----------------------------------------------------------------------------
where $\bm{\mathcal{G}}(E,\Phi) = \left[z^{+}{\bm I} - 
\bm{\mathcal{H}} \right]^{-1}$ 
is the Green's function of the system with 
$z^{+} = E + i\eta\ (\eta \rightarrow 0^{+})$, 
$\mathcal{N}$ is the number of sites in the lattice, 
and `$\textrm{Tr}$' represents the trace of the 
Green's function $\bm{\mathcal{G}}$. 
The results are displayed in Fig.~\ref{fig:dos}. For $\lambda=0$, 
we find that, the ADOS spectrum is not very densely populated 
[Fig.~\ref{fig:dos}(a)]; whereas for a non-zero value of $\lambda=0.9$, 
the ADOS spectrum becomes very densely packed [Fig.~\ref{fig:dos}(b)]. 
This indicates that, with a non-zero value of the hierarchy parameter 
$\lambda$, one can create more number of allowed states for the electrons 
to live in such a deterministic fractal geometry. This suggests towards 
a possible mechanism to engineer quantum states in a SPG fractal geometry. 
It is to be noted that, as our model involves a very complex configuration, 
therefore fabricating and realizing a large system in a real-life experiment 
in the laboratory may be a nontrivial task. We append the results for the 5th 
generation system in Appendix~\ref{appendix-1}.

It is important to see the manifestation of the above properties in terms 
of some experimentally measurable quantities. This brings us to the next 
section, where we demonstrate how one can control the persistent current 
as a function of the hierarchy parameter in this closed-loop fractal 
geometry. 
%%%%%%%%%%%%%%%%%%%%%%%%%%%%%%%%%%%%%%%%%%%%%%%%%%%%%%%%%%%%%%%%%%%%%%%
\section{Controlling the persistent current}
\label{sec:persistent-current} 
In isolated mesoscopic metallic rings threaded by a magnetic flux, one can 
have a dissipationless circulating current termed as persistent 
current~\cite{buttiker-PLA1983,gefen-PRL1989}. It is important to 
emphasize the fact that, this current has nothing to do with the persistent 
current found in a superconducting material. This idea was initially proposed 
for a single mesoscopic conducting loop structure~\cite{gefen-PRL1989, 
georges-bookchapter2007}, but later on people have also investigated the 
behavior of such persistent current in multi-loop fractal 
geometries~\cite{maiti-PRB2010}. The persistent current as a function of 
the magnetic flux $\Phi$ is related to the energy eigenvalues (energy 
levels) $E_{\alpha}$ at absolute zero temperature ($T=0$ K) as follows: 
%-----------------------------------------------------------------------------
\begin{equation} 
I_{\alpha}(\Phi) = -c\; \dfrac{\partial E_{\alpha}(\Phi)}{\partial \Phi},
\label{eq:persistent-current}
\end{equation}
%-----------------------------------------------------------------------------
where $c$ is a constant which is set equal to 1 for our calculation. 
We note that, for our model, we have the hierarchical 
distribution of the magnetic fluxes as $\Phi_{\ell}=\lambda^{\ell}\Phi$, 
where $\lambda$ is the hierarchy parameter, $\ell$ is the generation index, 
and $\Phi$ is some fundamental flux which can take values from 
$-\Phi_{0}$ to $\Phi_{0}$. Therefore, $\Phi$ in 
Eq.~\eqref{eq:persistent-current}, for our model can be expressed in terms of 
various orders of $\Phi_{\ell}$ and $\lambda^{\ell}$ depending on any given 
`$\ell$'-th generation fractal system. 
The total persistent current in the system is given by, 
%-----------------------------------------------------------------------------
\begin{equation} 
I(\Phi) = \sum_{\alpha=1}^{\mathcal{N}_{e}}I_{\alpha} = 
- \sum_{\alpha=1}^{\mathcal{N}_{e}} \dfrac{\partial E_{\alpha}}{\partial \Phi},
\label{eq:total-persistent-current}
\end{equation}
%-----------------------------------------------------------------------------       
where $\mathcal{N}_{e}$ is the total number of electrons in the system. We 
define $\nu = \mathcal{N}_{e}/\mathcal{N}(\ell)$ as the ``filling factor" 
for a $\ell$-th generation SPG fractal system. Note that, we consider 
non-interacting spinless electron model. 

\begin{figure}[ht]
\includegraphics[clip,width=0.49\columnwidth]{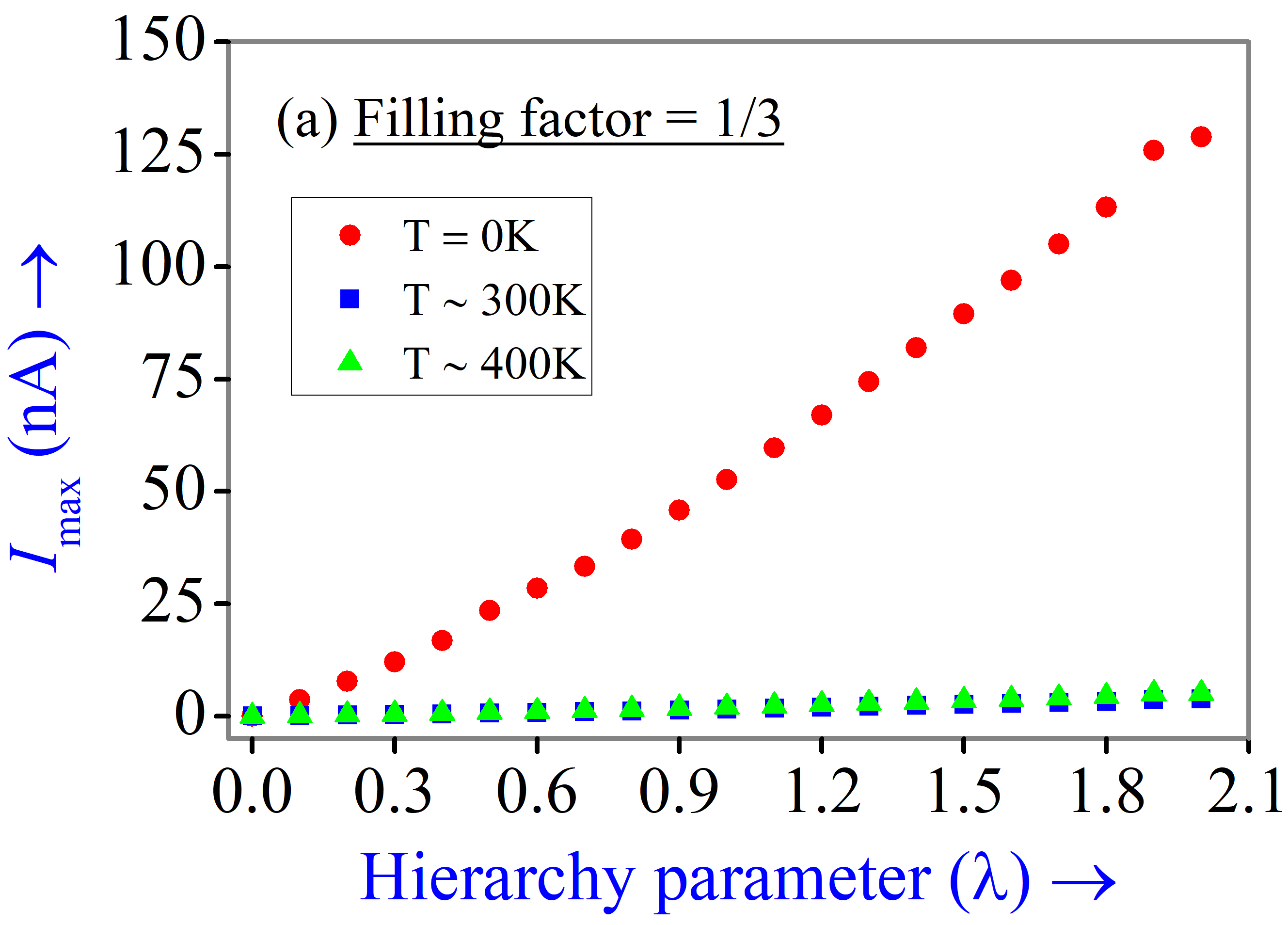}
\includegraphics[clip,width=0.49\columnwidth]{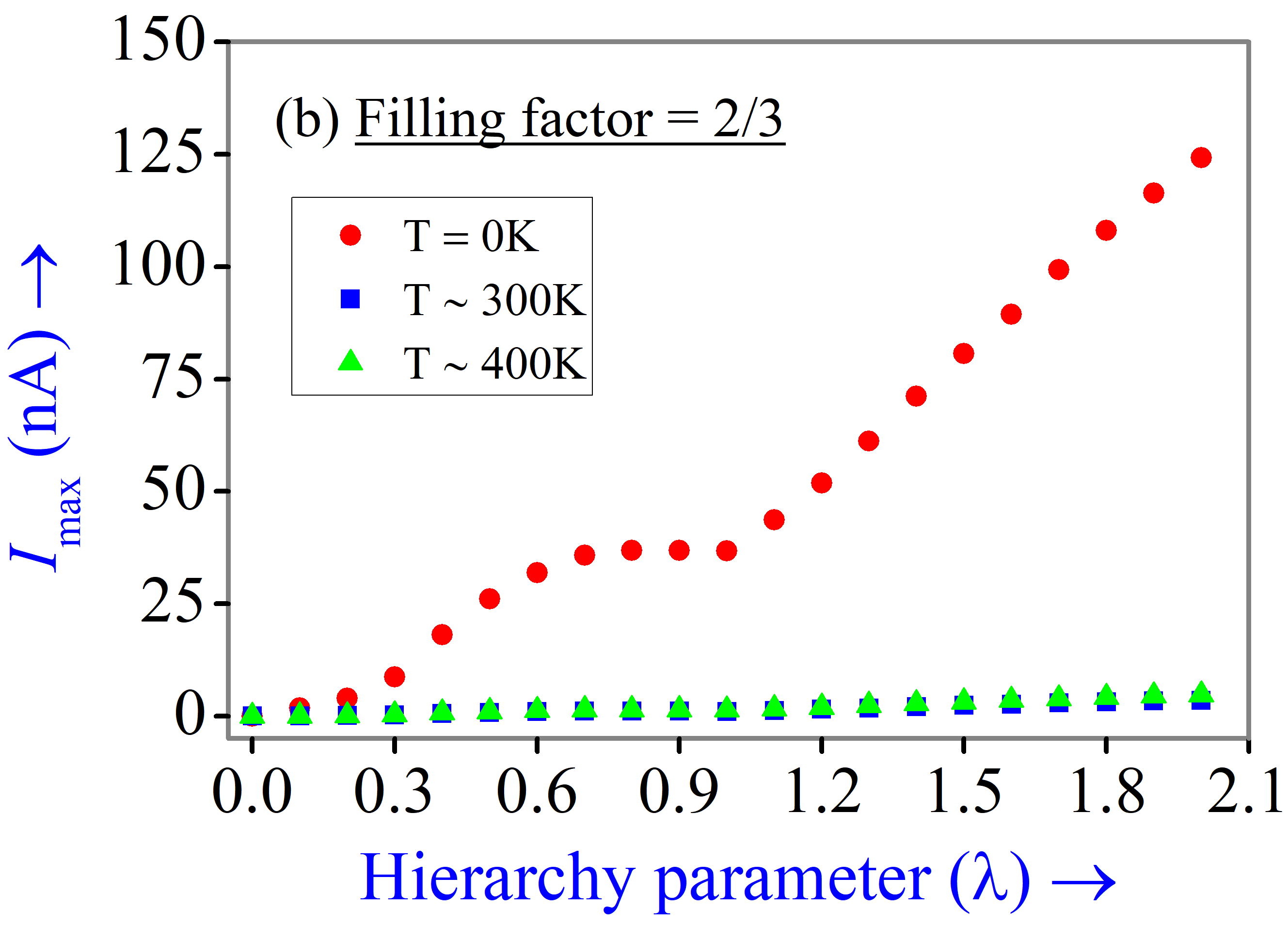}
\caption{Variation of the persistent current 
($I_{\textrm{max}}$) as a function of the hierarchy parameter 
$\lambda$ for a 4th generation SPG fractal structure with 
(a) $\nu=1/3$ and (b) $\nu=2/3$.}
\label{fig:persistent-current-gen4}
\end{figure}

For our model, the magnetic flux $\Phi$ depends on the hierarchy parameter 
$\lambda$. So, we can study the magnetic response of the SPG fractal geometry 
in terms of the persistent current as a function of $\lambda$. We have 
systematically investigated the behavior of the persistent current in a 
3rd generation SPG fractal network as a function of the hierarchy parameter 
$\lambda$ for different values of $\nu$. This is illustrated in 
Fig.~\ref{fig:persistent-current-gen3}(a)-(c). 
It is observed that, for different filling factors $\nu$, the persistent 
current in the system gets enhanced as one increases the value of $\lambda$. 
This opens up an interesting possibility by which one can flexibly 
control the persistent current in such multi-loop fractal geometries. 
Once again, this feature is also tested for a 4th generation SPG fractal 
structure (Fig.~\ref{fig:persistent-current-gen4}), and found to be true 
as an evidence of the generic self-similar property of a fractal structure. 
We remark that, in a real experimental setup, if one typically chooses the 
system size to be $\sim \mu$m, magnetic field $\sim 10$ Tesla, then the persistent 
current we get will be $\sim$ nano-Amp (nA). 
%######################################################
%###################################################### 

It is important to briefly mention the effect of the finite 
temperature on the persistent current for our fractal lattice system. 
At finite temperature ($T>0$), one can calculate the persistent current in the system 
using the following formula~\cite{bloch-PRB1970,wendler-PSSB1995}: 
%-----------------------------------------------------------------------------
\begin{subequations}
\begin{align}
& I(\Phi) = - \dfrac{\partial F(\Phi)}{\partial \Phi} \\
\text{with } & F(\Phi) = -k_{B}T\:\ln\Big(\sum_{\nu} -\dfrac{E_{\nu}}{k_{B}T} \Big)
\label{eq:total-persistent-current-in-finite-temp}
\end{align} 
\end{subequations}
%-----------------------------------------------------------------------------
where $F$ is the Helmholtz free energy, $k_{B}$ is the Boltzmann constant, and 
$T$ is the absolute temperature. 
It is found that, at finite temperature, the persistent 
current in the system decreases drastically as evident from Fig.~\ref{fig:persistent-current-gen3} 
and Fig.~\ref{fig:persistent-current-gen4}. This is because, as we go to the finite temperature, 
electrons can occupy the higher energy levels with different phases, which tend to 
cancel out the net current in the system.  
%######################################################
\begin{figure}[ht]
\includegraphics[clip,width=0.7\columnwidth]{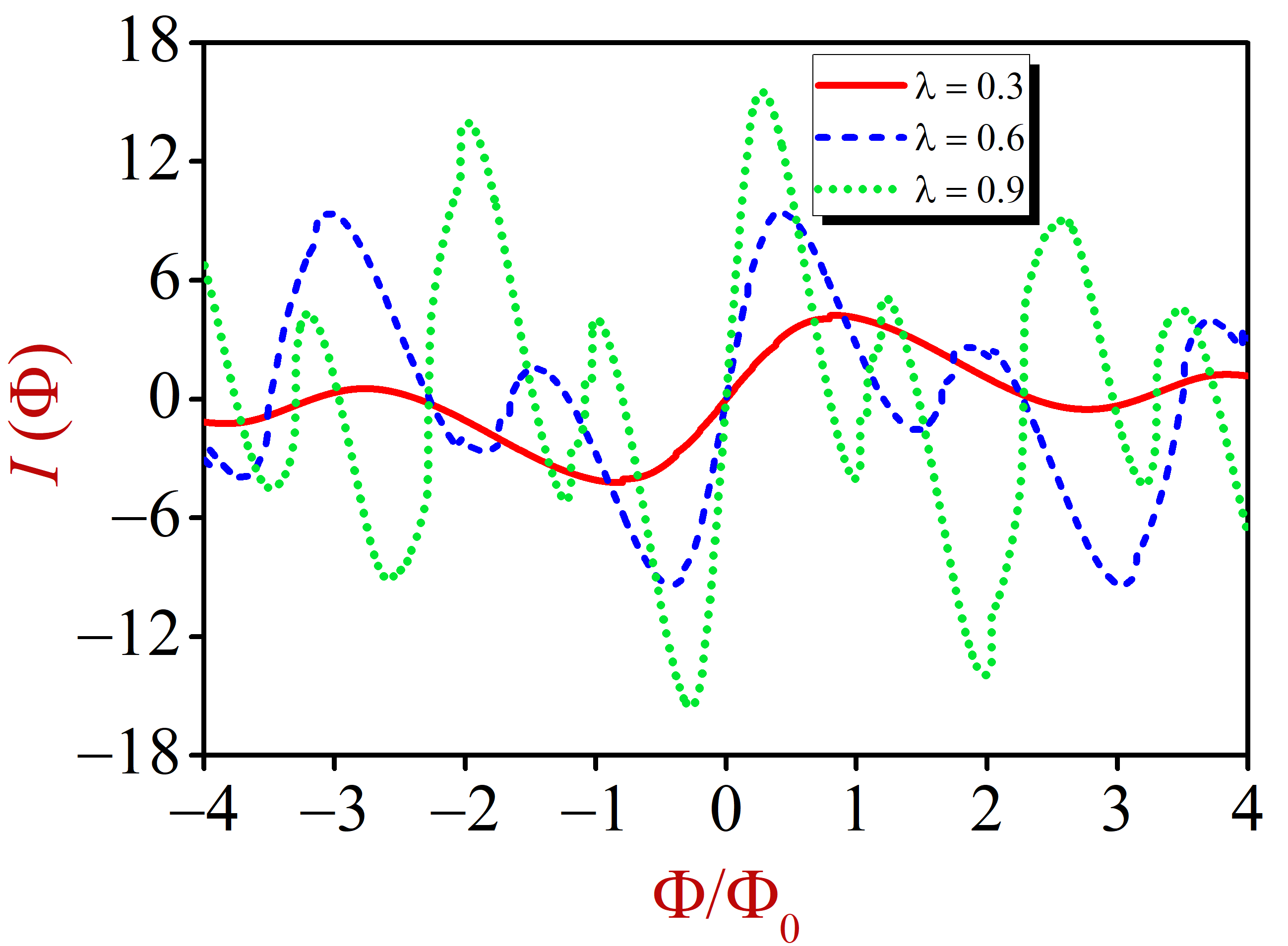}
\caption{Variation of the persistent current ($I$) as a function of the 
magnetic flux for three different values of the hierarchy parameter 
$\lambda=0.3$ (red solid line), $\lambda=0.6$ (blue dashed line), and 
$\lambda=0.9$ (green dotted line), respectively in a $3$rd generation 
SPG fractal structure. The filling factor is set be $\nu=1/3$.}
\label{fig:persistent-current-vs-flux}
\end{figure}
%######################################################  

The variation of the persistent current with the magnetic flux for different fixed values 
of the hierarchy parameter $\lambda$ is shown in Fig.~\ref{fig:persistent-current-vs-flux}. 
From these plots, we can clearly identify that as the value of $\lambda$ increases, 
the magnitude of the persistent current in the system also increases. This result is 
consistent with the result shown in Fig.~\ref{fig:persistent-current-gen3}. We note that, the 
periodicity of the persistent current as a function of the magnetic flux also increases as 
we increase the value of $\lambda$. This is because of the hierarchical nature of the magnetic 
flux distribution in the system as a function of the hierarchy parameter $\lambda$.

It is to be noted that, in principle, one could study these properties 
for any other higher generation SPG fractal structures. We predict that, 
our results will be qualitatively retained for those cases also. 
However, from the experimental point-of-view, one cannot go beyond 
certain system size in the real-life laboratory when it comes to 
fabricating such complex fractal geometries.  
%%%%%%%%%%%%%%%%%%%%%%%%%%%%%%%%%%%%%%%%%%%%%%%%%%%%%%%%%%%%%%%%%%%%%%%
\section{Summary and future outlook}
\label{sec:summary} 
In conclusion, we have studied the role of a hierarchical distribution 
of magnetic flux on the energy spectrum and persistent current in a 
multi-loop Sierpinski gasket fractal structure. We have shown that, by 
controlling the hierarchy parameter of the embedded magnetic flux in 
this SPG fractal network, one can effectively engineer quantum states 
in such a complex system. From the application point-of-view, we have 
shown that this directly impacts the behavior of the persistent current 
in this SPG fractal structure. We can suitably control the persistent 
current in the system by tunning the hierarchy parameter in a flexible 
way for different filling factors of the electronic states. One could 
in principle, test these results by designing an experiment using a 
self-assembled molecular SPG fractal structure. This could be helpful 
to devise an unconventional nanoelectronic device using a molecular SPG 
fractal structure. In future, we plan to generalize this idea for other 
similar kind of closed-loop fractal geometries, such as 
Vicsek fractal~\cite{pal-PRB2012,pal-PRA2013}. 
Another interesting future prospect of this work could be to study these 
properties in fractal lattices with self-affine feature~\cite{self-affine}.    
%%%%%%%%%%%%%%%%%%%%%%%%%%%%%%%%%%%%%%%%%%%%%%%%%%%%%%%%%%%%%%%%%%%%%%%
%*********************************************************
\begin{acknowledgments}
The author would like to thank Prof.\ Arunava Chakrabarti from Presidency 
University, Kolkata for his useful feedback on the initial stage of this 
work. The author also acknowledges helpful discussion with Dr.\ Amrita 
Mukherjee from TIFR, Mumbai on the calculation of the persistent current.  
\end{acknowledgments}
%*********************************************************

%*********************************************************
\section*{Data Availability Statement}
The data that support the findings of this study are available from the 
corresponding author upon reasonable request.
%*********************************************************
%%%%%%%%%%%%%%%%%%%%%%%%%%%%%%%%%%%%%%%%%%%%%%%%%%%%%%%%%%%%%%%%%%%%%%%
\appendix*
%-----------------------------------------------------------------------------------
%%%%%%%%%%%%%%%%%%%% APPENDIX %%%%%%%%%%%%%%%%%%%%
\section{Results for higher generation system}
\label{appendix-1}
\begin{figure}[ht]
\centering
\includegraphics[clip, width=0.9\columnwidth]{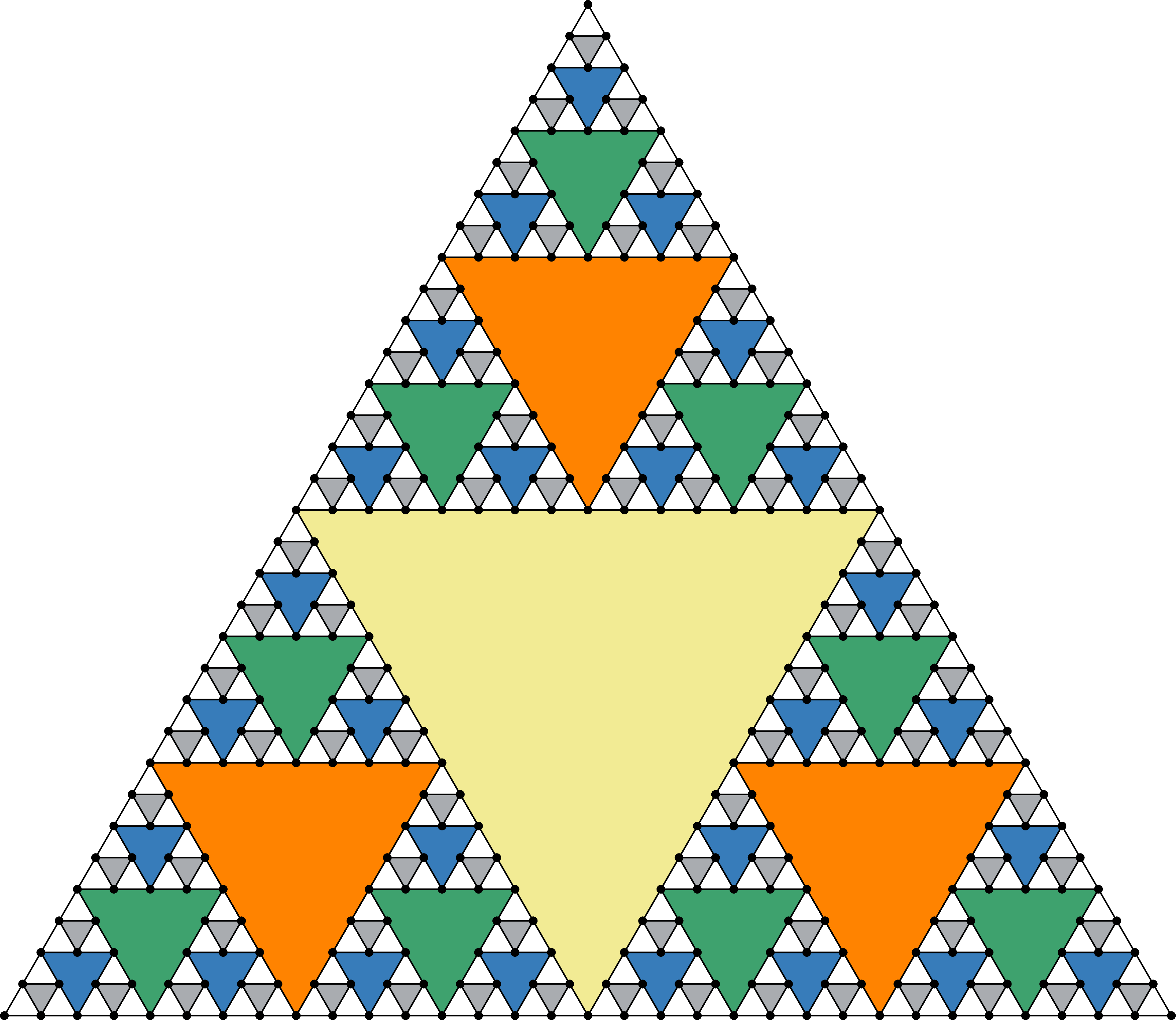}
\caption{Schematic representation of the 
$5$th generation SPG fractal lattice geometry with hierarchical 
distribution of magnetic fluxes in the downward triangular plaquettes. 
Different colors represent the hierarchical flux distribution 
according to the order of hierarchy of the downward triangles (from 
smallest to the largest), \viz $\Phi_{1}$, $\Phi_{2}$, $\Phi_{3}$, $\Phi_{4}$, 
and $\Phi_{5}$.}
\label{fig:Scheme-for-Hamiltonian}
\end{figure}

\begin{figure}[ht]
\includegraphics[clip,width=0.49\columnwidth]{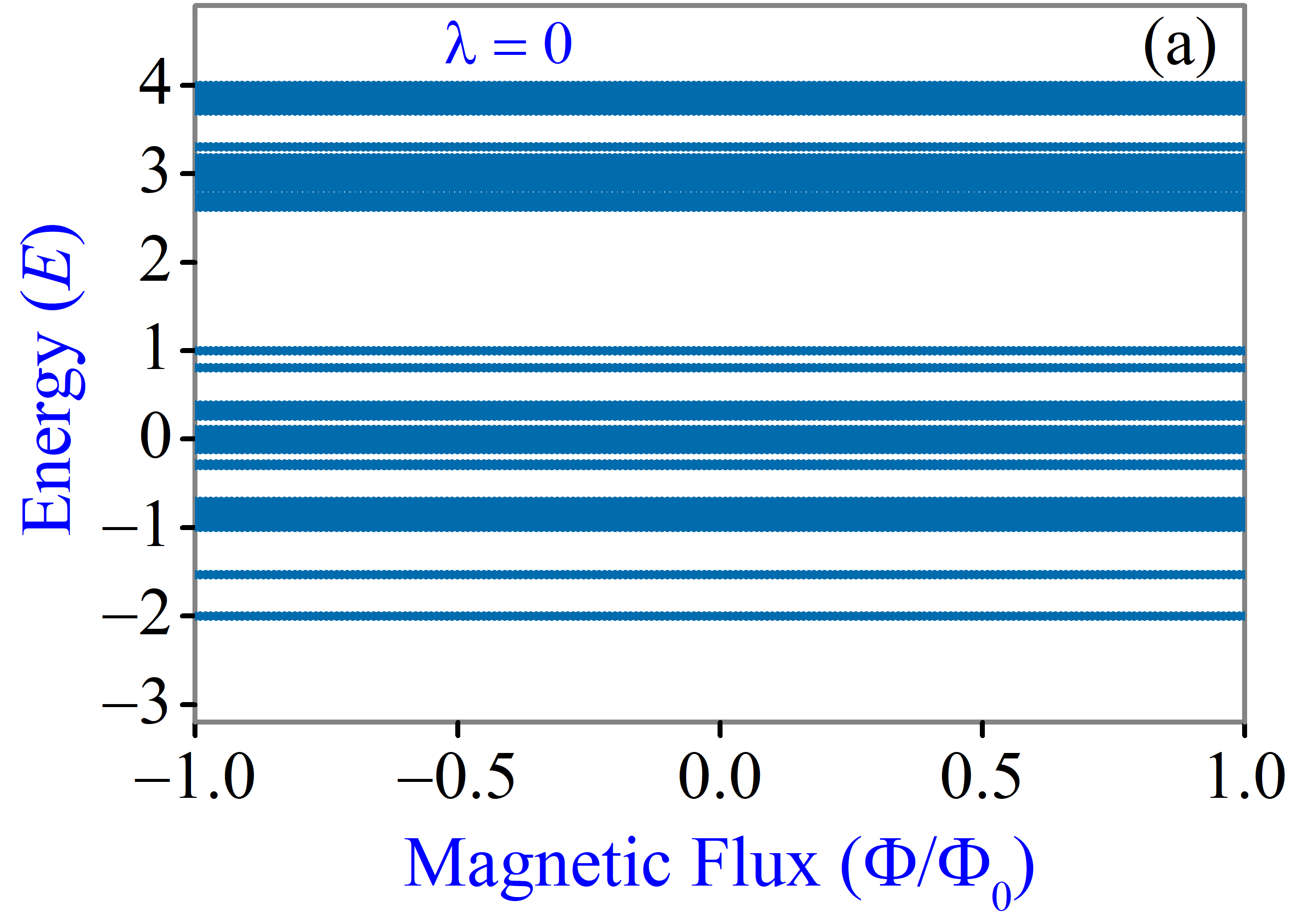}
\includegraphics[clip,width=0.49\columnwidth]{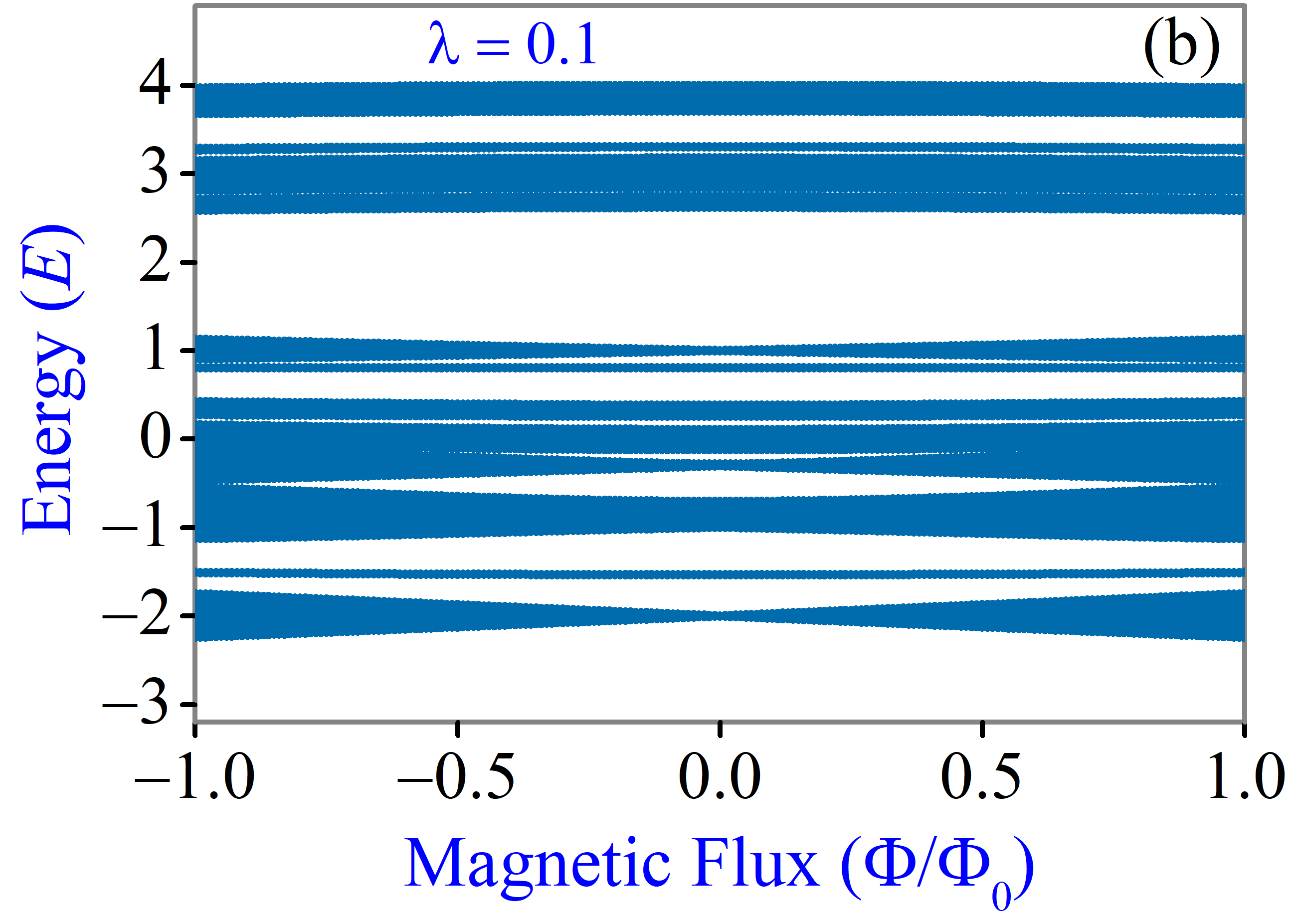}
\vskip 0.2cm
\includegraphics[clip,width=0.49\columnwidth]{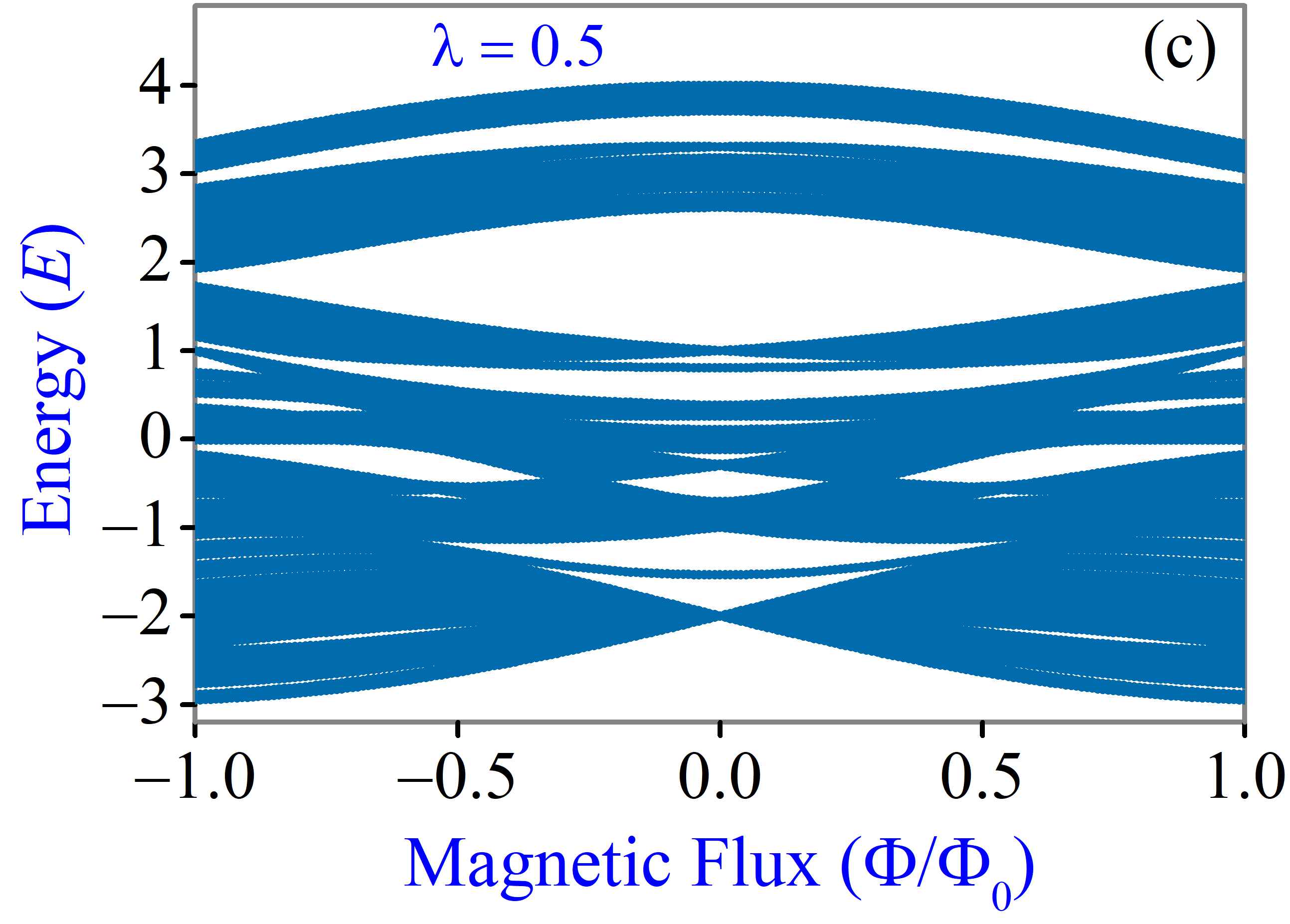}
\includegraphics[clip,width=0.49\columnwidth]{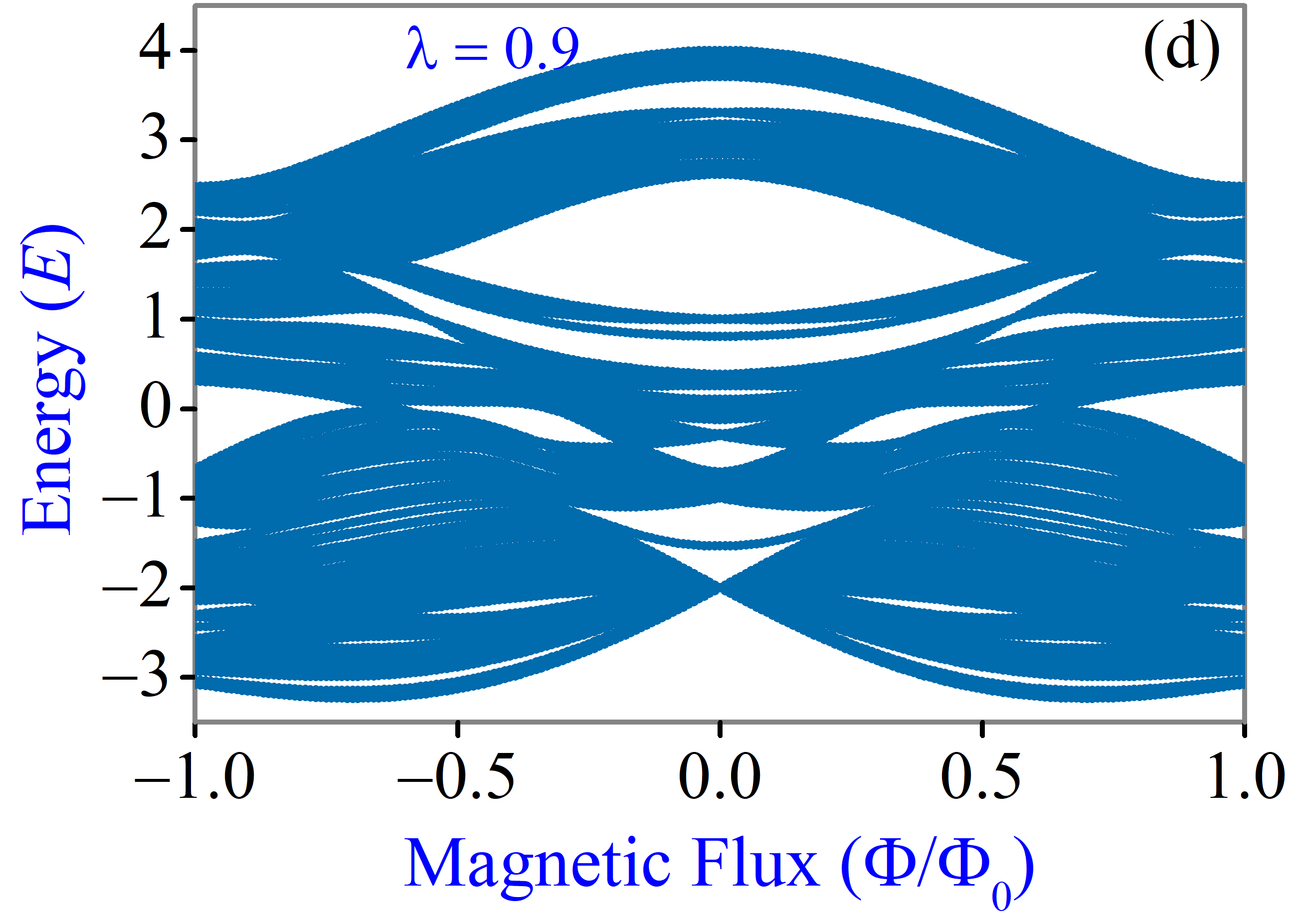}
\caption{Energy eigenvalue spectrum for a 5th 
generation SPG lattice structure (having $\mathcal{N}(5) = 366$ sites) 
as a function of the hierarchical magnetic flux with 
values of the hierarchy parameter chosen to be same as 
in Fig.~\ref{fig:eng-spec-gen3} and Fig.~\ref{fig:eng-spec-gen4}.}
\label{fig:eng-spec-gen5}
\end{figure}

We remark that, as our model demand a very complicated implementation scheme in an actual 
real-life experimental situation, it could be difficult to fabricate and do the experiment 
for a large system. However, to convince the readers, in this appendix, we show the results 
for a 5th generation system (Fig.~\ref{fig:Scheme-for-Hamiltonian}) with 
$\mathcal{N}(5) = 366$ sites as displayed 
in Fig.~\ref{fig:eng-spec-gen5}. 
%
%------------------- Appendix Figure 1 ---------------------
%@@@@@@@@@@@@@@@@@@@@@@@@@@@@@@@@@@@@
%\renewcommand\thefigure{A.\arabic{figure}}    
%\setcounter{figure}{0} 
%@@@@@@@@@@@@@@@@@@@@@@@@@@@@@@@@@@@@  
%
%------------------- Appendix Figure 2 ---------------------
%@@@@@@@@@@@@@@@@@@@@@@@@@@@@@@@@@@@@ 
%@@@@@@@@@@@@@@@@@@@@@@@@@@@@@@@@@@@@ 

We found that, all our predictions from the previous generation systems (discussed in the 
main text) holds good for this relatively bigger system size also. It is worth mentioning 
that, as we have a deterministic disorder distribution of AB phases along different bonds 
for any given generation for our lattice model, there is no ready-made algorithm to generate 
the Hamiltonian for the system -- one has do it by hand for each given generation.
%------------------- Appendix Figure 3 ---------------------
%@@@@@@@@@@@@@@@@@@@@@@@@@@@@@@@@@@@@ 
\begin{figure}[ht]
\includegraphics[clip,width=0.49\columnwidth]{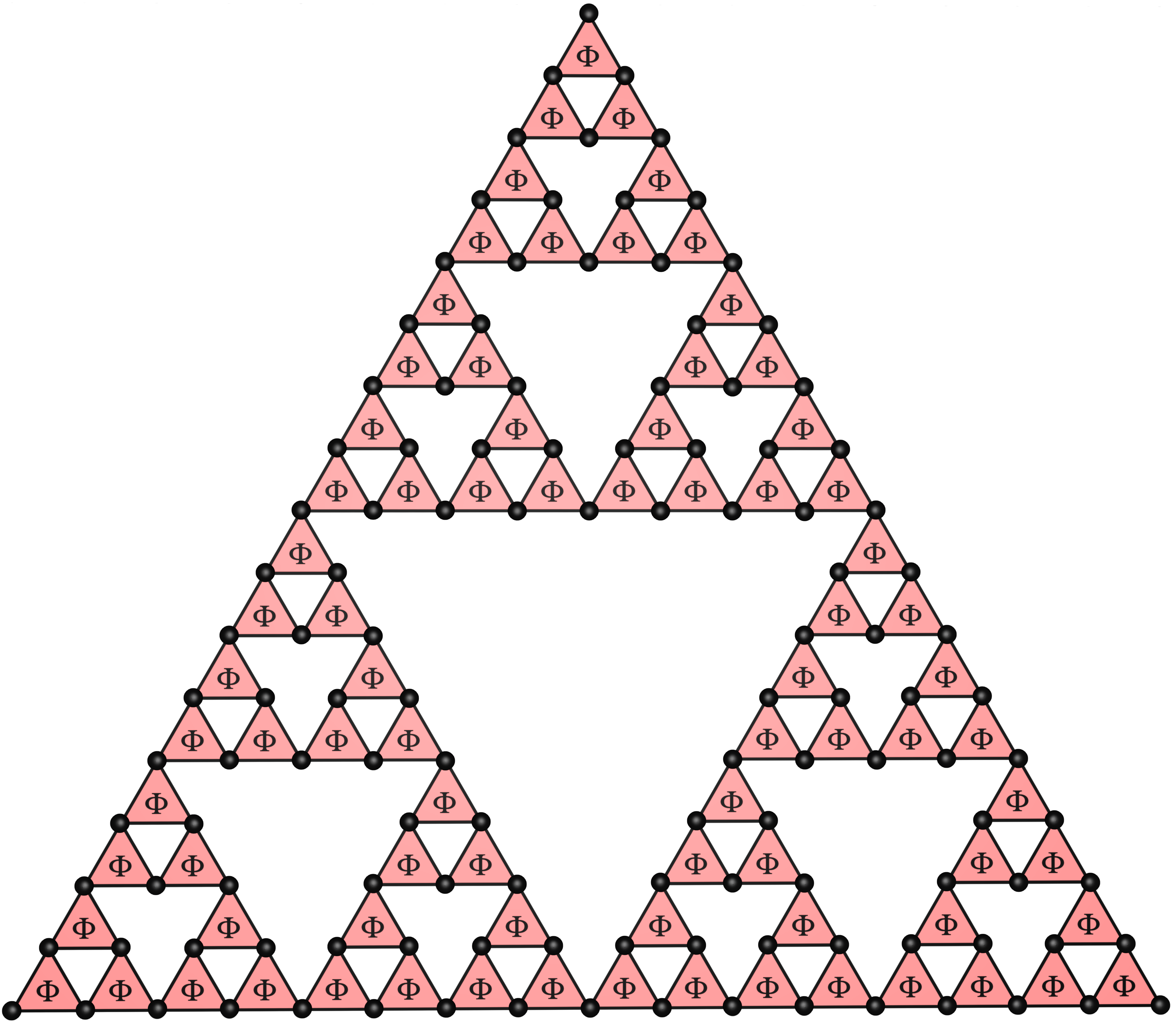}
\includegraphics[clip,width=0.49\columnwidth]{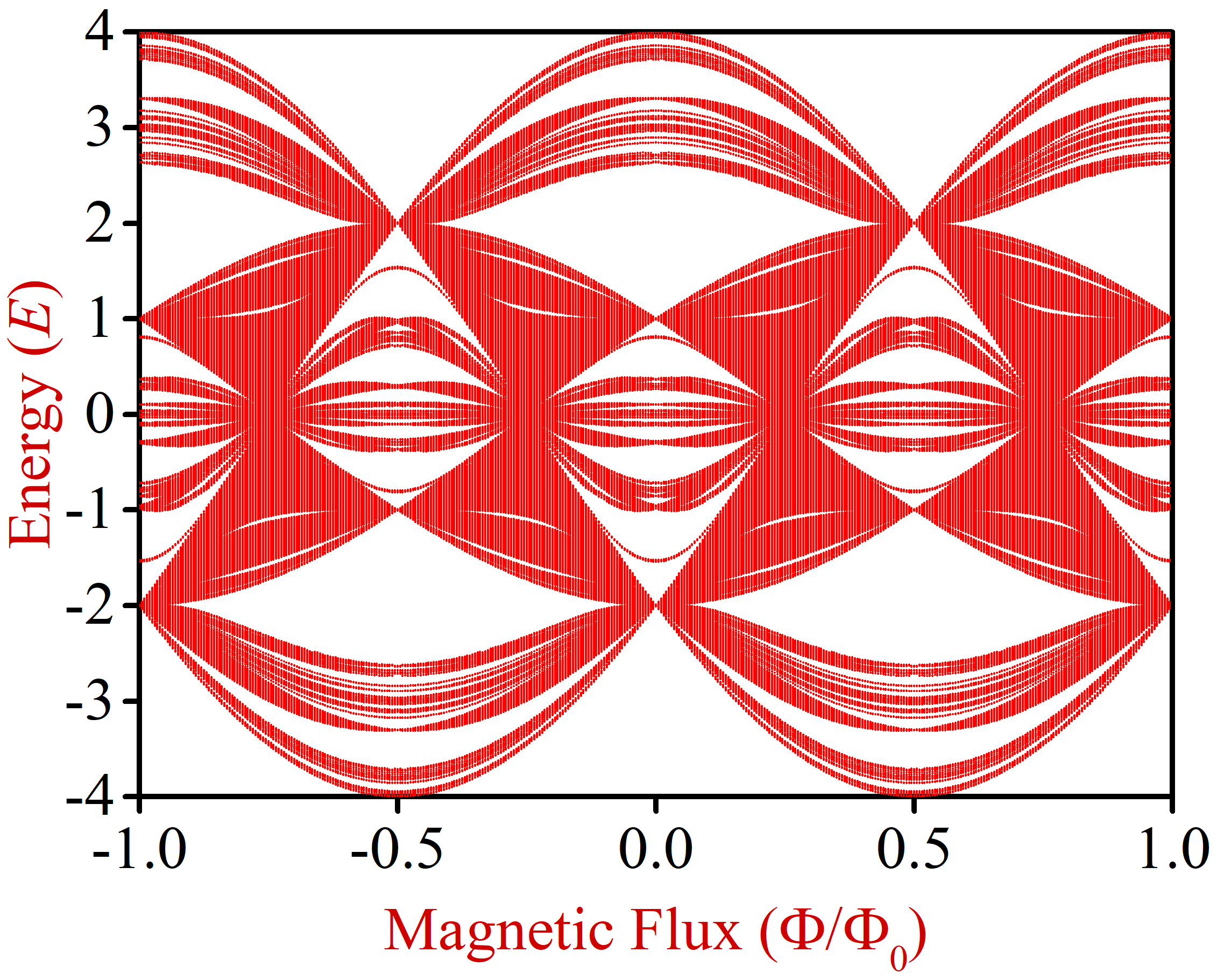}
\caption{Schematic representation of a SPG fractal structure with a uniform 
distribution of the magnetic flux in all the upward triangular loops (left panel) 
and its energy eigenvalue distribution (right panel).}
\label{fig:spg-with-uniform-flux}
\end{figure}
%@@@@@@@@@@@@@@@@@@@@@@@@@@@@@@@@@@@@  

In this study, we have considered the distribution of the magnetic flux 
in the downward triangular loops of the SPG fractal structure in such a way that it follows 
a special kind of hierarchical pattern. However, it is also interesting 
see what will happen if we put the magnetic flux in all the upward triangular loops of the SPG 
structure. This will impose a uniform Aharonov-Bohm phase factor along all the bonds in the SPG 
fractal structure (see Fig.~\ref{fig:spg-with-uniform-flux}). In this case, there will be no 
concept of hierarchical flux distribution as automatically imposed by the ramification pattern 
of the SPG fractal geometry. We have done the computation for this case also and the results 
for the energy eigenvalue spectrum for a finite generation SPG fractal structure is shown in 
Fig.~\ref{fig:spg-with-uniform-flux}.
%-----------------------------------------------------------------------------------

\end{document}